# White Paper on Phototrophic Biosignatures: Research Priorities for the Search for Life on Other Worlds

**A product of the Nexus for Exoplanet System Science**
**2026 Extraterrestrial Photosynthesis Workshop**
**https://nexss.info/extraterrestrial-photosynthesis-workshop/**

**Authors:**

Nancy Y. Kiang, NASA Goddard Institute for Space Studies, New York, New York, USA, nancy.y.kiang@nasa.gov

Christopher Gisriel, University of Wisconsin, Madison, Wisconsin, USA, gisriel@wisc.edu

Robert E. Blankenship, Emeritus, Arizona State University, Washington University in St. Louis, reblankenship@gmail.com

Mary N. Parenteau, NASA Ames Research Center, Moffett Field, California, USA, mary.n.parenteau@nasa.gov

Hazel A. Barton, University of Alabama, Tuscaloosa, Alabama, United States. hbarton1@ua.edu

Oded Béjà, Technion–Israel Institute of Technology, Haifa, Israel, beja@technion.ac.il

Anthony J. Burnetti, Georgia Institute of Technology, Atlanta, Georgia, United States. anthony.burnetti@biosci.gatech.edu

Ligia F. Coelho, Division of Geological and Planetary Sciences, California Institute of Technology, Pasadena, CA, USA, lc992@caltech.edu

Saleheh Ebadirad, Department of Earth and Planetary Sciences, University of California, Riverside, CA, USA, saleheh.ebadirad@email.ucr.edu

Nathan M. Ennist, Department of Molecular and Structural Biochemistry, North Carolina State University, Raleigh, NC, USA, nmennist@ncsu.edu

Yuka Fujii, Division of Science, National Astronomical Observatory of Japan, Tokyo, Japan, yuka.fujii.ebihara@gmail.com

Colin Gates, Loyola University Chicago, Chicago, Illinois, USA, cgates4@luc.edu

R.J. Graham, Department of Geophysical Sciences, University of Chicago, Chicago, IL, USA arejaygraham@uchicago.edu

Keiichi Inoue, The University of Tokyo, Kashiwa, Chiba, Japan, kinoue@edu.k.u-tokyo.ac.jp

Betül Kaçar, Department of Bacteriology, University of Wisconsin, Madison, Wisconsin, USA, bkacar@wisc.edu

Yu Komatsu, Graduate School of Science and Engineering, Ibaraki University, Hitachi, Ibaraki, Japan, yu.komatsu.za71@vc.ibaraki.ac.jp

Émilie A. Laflèche, Department of Earth, Atmospheric and Planetary Sciences, Purdue University, West Lafayette, IN, USA, elaflech@purdue.edu

Nicoletta La Rocca, Department of Biology, University of Padova, Padova, Italy, nicoletta.larocca@unipd.it

Elisabetta Liistro, Department of Biology, University of Padova, Padova, Italy, elisabetta.liistro@phd.unipd.it
Jonathan Lindsey, Dept. of Chemistry, North Carolina State University, Raleigh, North Carolina, USA, jlindsey@ncsu.edu
Timothy Lyons, Department of Earth and Planetary Sciences, University of California, Riverside, California, USA, timothyl@ucr.edu
Carolina A. Martinez-Gutierrez, University of California Santa Barbara, Santa Barbara, CA, USA, cmartinezg@ucsb.edu
Taro Matsuo, Department of Earth and Space Science, Graduate School of Science, University of Osaka, Osaka Japan, matsuo@ess.sci.osaka-u.ac.jp
Victoria Meadows, SETI Institute, Mountain View, California, USA, vmeadows@seti.org
Gary F. Moore, School of Molecular Sciences and the Biodesign Institute Center for Applied Structural Discovery (CASD), Arizona State University, Tempe, Arizona, USA, gary.f.moore@asu.edu
Massimo Olivucci, Department of Biotechnology, Chemistry and Pharmacy, University of Siena, Italy & Department of Chemistry, Bowling Green State University, OH 43403, USA olivucci@unisi.it & molivuc@bgsu.edu
Kevin E. Redding, School of Molecular Sciences, Arizona State University, Tempe, Arizona, USA, Kevin.Redding@asu.edu
Edward W. Schwieterman, Department of Earth and Planetary Sciences, University of California Riverside, Riverside, California, USA, eschwiet@ucr.edu
Tejinder Singh, NASA Ames Research Center, Moffett Field, California, USA, tejinder.singh@nasa.gov
Kenji Takizawa, Astrobiology Center, Okazaki, Aichi, Japan, kenji-t@nibb.ac.jp
Anna Grace Ulses, Department of Earth & Space Sciences, University of Washington, WA, USA aulses@uw.edu
Junko Yano, Lawrence Berkeley National Laboratory, Berkeley, California, USA JYano@lbl.gov
Felisa Wolfe-Simon, Bay Area Environmental Research Institute, Moffett Field, California, USA, felisawolfesimon@gmail.com
Michael L. Wong, Earth & Planets Laboratory, Carnegie Institution for Science, Washington, DC 20015, USA, mwong@carnegiescience.edu
Anastasia G. Yanchilina, SETI Institute, Mountain View, California, USA, ayanchilina@seti.org

**Endorsements:**

**Full name, Institution, City, State, Country, email address.**

Karl R. Stapelfeldt, NASA Exoplanet Exploration Program Office, Mail Stop 301-320, Jet Propulsion Laboratory, 4800 Oak Grove Drive, Pasadena CA 91109 USA, karl.r.stapelfeldt@jpl.nasa.gov
Tolga Korkut, Rutgers, The State University of New Jersey, New Brunswick, NJ, USA tk742@scarletmail.rutgers.edu
Simon Fronmueller, University of Minnesota Twin Cities, Saint Paul, MN, USA fronm002@umn.edu

**White Paper on Phototrophic Biosignatures:**
**Research Priorities for the Search for Life on Other Worlds**

## Executive summary

Photosynthesis is of prime interest in the telescopic search for life beyond the Solar System, because, as demonstrated on Earth, oxygenic photosynthesis has the capacity to dramatically transform the chemistry of a planet and produces two strong "biosignatures" as signs of life that can be seen from space at the global scale: atmospheric oxygen and the Vegetation Red Edge (VRE). The VRE is the reflectance signature of plant leaves, characterized by a step-like increase in reflectance from the red to the near-infrared. The absorption in the red is due to chlorophyll *a* (Chl *a*), the primary photopigment of oxygenic photosynthesis, responsible for both the color of the VRE as a surface biosignature and for the charge separation that allows water oxidation and hence the production of oxygen. While Chl *a* dominates our planet, the Earth harbors a wider variety of phototrophic organisms with other pigments that produce edge-like spectral features across the UV/VIS/NIR adapted to niche environments, which are naturally suggestive of diverse signatures that could be found on other planets. Identifying and predicting edge-like reflectance features of surface biota that depart from the VRE have motivated many astrobiology studies seeking alternative surface biosignature candidates for scenarios of the Archean and Proterozoic Earth and for photosynthetic life adapted to other stars. However, the astrobiology community remains very much at an early stage in its ability to constrain the probability that an observation of another planet has detected a sign of photosynthetic life given that data in the context of the star-planet system. This white paper identifies critical questions for researchers to answer to advance the search for phototrophic biosignatures to a predictive capability, in collaborations across the wide array of scientific disciplines that provide the astrobiological context. These questions pertain to the origins, key features, diversity, and potential for alternative adaptations in fundamental components of phototrophy: light harvesting; the electron transfer pathway in photosynthesis; rhodopsin-based proton-pumping; and carbon fixation in autotrophy. We also identify research needs to constrain how these molecular mechanisms evolve and compete, and how evolution and ecology affect how they scale through time over a planet to have the potential to be detected by a direct imaging mission. The research questions and recommendations identified here are cross-linked to those posed by the last NASA Astrobiology Strategy 2015, and they are classified into the Focus Areas of the upcoming NASA Decadal Astrobiology Exploration Strategy (DARES). We hope this white paper will inspire more scientists to probe the hard questions about searching for phototrophic signs of life and provide the impetus for advancing the Astrobiology community around this now pressing precursor science for the Habitable Worlds Observatory.

# White Paper on Phototrophic Biosignatures:
# Research Priorities for the Search for Life on Other Worlds

## Table of Contents

## 1. Introduction

Photosynthesis is of primary interest in the telescopic search for life beyond the Solar System, because oxygenic photosynthesis has the capacity to dramatically transform the chemistry of a planet (as demonstrated on Earth), and produces two strong "biosignatures" as signs of life that can be seen from space at the global scale: atmospheric oxygen and the Vegetation Red Edge (VRE) (Des Marais et al., 2002; Des Marais et al., 2008; Meadows et al., 2018; Sagan et al., 1993). Photosynthesis, both anoxygenic and oxygenic, emerged early in our planet's history(Oliver et al., 2023; Tice & Lowe, 2004; Westall et al., 2006), allowing life to capture the stellar radiation of our Sun as an ubiquitous energy source, enough eventually to support a global biosphere. Therefore, the exoplanet and astrobiology communities anticipate that photosynthesis could emerge and manifest in planetary-scale biosignatures detectable to upcoming telescope technologies (LUVOIR & Team, 2019)(Habitable Worlds Observatory Living Worlds Working Group Science Case Document, 2025, (HWO25, 2025)). Direct imaging telescopes like the developing Habitable Worlds Observatory (HWO) concept mission are slated to obtain reflected light from planets around Sun-like stars. This potential to characterize surface properties of planets has renewed interest in surface biosignatures (Parenteau et al., 2025), and makes it pressing now to develop measurement requirements and methods to interpret the data for evidence of surface life to potentially enhance the science return of the mission. However, the astrobiology community remains very much at an early stage in its ability to constrain the probability that an observation of another planet has detected a sign of photosynthetic life given that data in the context of the star-planet system.

The VRE is the reflectance signature of plant leaves (weakly present in cyanobacteria and algae), characterized by a step-like increase in reflectance from the red to the near-infrared (Gates et al., 1965) . It is detectable in satellite (Tucker & Maxell, 1976), spacecraft (Sagan et al., 1993), Earthshine (Woolf et al., 2002), and simulated telescope observations (Seager et al., 2005) observations of the Earth. Spanning ~680 to ~760 nm, the feature is due to the tailing off of absorbance by Chl a (the "red drop", (Emerson & Lewis, 1943)) and the high scattering in the NIR is due to the change in index of refraction between leaf mesophyll cells walls and air spaces inside the leaf (Gausman et al., 1974). Its exact position and steepness can vary due to tuning of the chlorophylls in the core antenna, ratio of Photosystems I and II (PSI, PSII), and the presence of other pigments in different organisms, a subject of ongoing research by the Earth Science and photosynthesis research communities. Identifying and predicting edge-like reflectance features of surface biota that depart from the Vegetation Red Edge (VRE) have motivated many astrobiology studies seeking alternative surface biosignature candidates for scenarios of the Archean and Proterozoic Earth and for photosynthetic life adapted to other stars, particularly M star with low visible and high NIR light flux. There remains much yet to be understood to achieve a predictive capability.

This white paper summarizes research questions to advance the search for alien phototrophic life identified at the 2026 Nexus for Exoplanet System Science (NExSS) Extraterrestrial Photosynthesis Workshop held in May 2026. The scope of this paper covers phototrophy in general as a potential producer of biosignatures relevant to the search for life on exoplanets. Whereas "photosynthesis" usually refers specifically to light use for autotrophy, "phototrophy" more broadly means the biological conversion of light energy into stored chemical energy, which includes rhodopsin-based systems. This paper is also focused largely on the search for life on exoplanets, while acknowledging the still open possibility for past phototrophy on Mars or even current phototrophy in the clouds of Venus. We summarize developments in the past two decades in both photosynthesis research and planetary/exoplanet science, as well as identify research gaps that will afford opportunities for scientists to advance the astrobiology of phototrophy. For both Earth and exoplanet contexts, this white paper organizes a summary of key unknowns about photosynthesis with regard to critical questions on origins of different components of the photosynthetic apparatus, co-evolution with the planet, and scaling across time and space.

## 2. Developments since the last Astrobiology decadal strategy reports

Astrobiology may be defined as “the study of the origin, evolution, distribution, and future of life in the universe” (National Academies of Sciences, 2019)("NASEM, 2019"). The last comprehensive decadal survey studies to map the next directions for astrobiology research were the *NASA Astrobiology Strategy 2015* (Hays et al., 2015) (“NASA2015”) and *Origins, Worlds, and Life: A Decadal Strategy for Planetary Science and Astrobiology 2023-2032* (National Academies of Sciences, 2023) (“OWL2023”). The explosion in exoplanet discoveries and their intertwining with astrobiology prompted targeted NASEM consensus reviews, on Exoplanet Science Strategy (National Academies of Sciences, 2018) (“NASEM2018Exo”) and An Astrobiology Strategy for the Search for Life in the Universe (National Academies of Sciences, 2019) (“NASEM2019Astrobio”). The inaugural open community conference for NASA’s Habitable Worlds Observatory (HWO), a flagship direct imaging space telescope mission, was held July 28–31, 2025, following many online working group meetings to develop Science Case Development Documents (SCDDs) with many precursor science studies (“HWO2025”). NASA is currently developing a synthesis study, the NASA Decadal Astrobiology Research and Exploration Strategy (NASA-DARES) of community perspectives on science gaps, emerging research, opportunities, needs, and connections across NASA Science to provide context for future planning activities.

NASA2015 marked a stage of increasing cross-disciplinary maturity in astrobiology research, and identified 6 major areas of research resonant in the field: 1) identifying abiotic sources of organic compounds; 2) synthesis and function of macromolecules in the origin of life; 3) early life and increasing complexity; 4) co-evolution of life and the physical environment; 5) identifying, exploring, and characterizing environments for habitability and biosignatures; and 6) constructing habitable worlds. It emphasized the need for astrobiology research to constrain whether evolutionary innovations would be inevitable versus unusual chance occurrences, given the environmental context. It focused on three classes of innovations: the evolution of biological complexity (starting with the cooperative coupling between biopolymers), the acquisition of energy, and, separately, phototrophy as transformative to the composition and energy budget of the Earth. The report broadly laid out these questions about phototrophy:

1. How soon after the origin of life did phototrophy arise?
2. How does phototrophy impact the way life interacts with the surrounding geochemistry, and to what extent does this relationship provide biosignatures?

3. What did the earliest forms of phototrophy look like?
4. What photoreactive chemicals were available in the environment of early life?
5. Was phototrophy adapted from simpler forms of light-driven energy transduction, UV protection mechanisms, trophic mechanisms, or some other process?
6. When did organisms evolve oxygenic photosynthesis (OP), what were its evolutionary precursors, and how is this innovation related to the general oxidation of the Earth's crust and atmosphere?
7. How much biologically useful energy is produced by microbial rhodopsins and other photoreactive biochemistries? Is this enough to sustain an organism?
8. Are there undiscovered types of phototrophy present on Earth today?
9. What types of light can be biologically useful and how does the range of phototrophy constrain our concept of habitable planets?
10. How will the biosignatures of pigments differ for different atmospheric compositions and parent stellar type?

NASA2015 also identified the recent methodological developments that empower advances in the field, through genomics, informatics, computational biology.

NASEM2019Astrobio, tasked with reviewing how mission priorities for Solar System and exoplanet exploration addressed the search for life, recognized astrobiology as inherently a system science integrating "interactions within and between the physical, chemical, biological, geologic, planetary, and astrophysical systems as they relate to understanding how an environment transforms from nonliving to living and how life and its host environment coevolve." The report incorporated the findings of NASEM2018Exo where exoplanet observations and science must constrain the star-planet context for life, highlighting concepts of dynamic habitability, comparative planetology and multiparameter habitability assessment, and augmenting the NASA2015 with the need to identify novel biosignatures (other than well-known oxygenic photosynthesis), including agnostic biosignatures that manifest not as specific molecules but global phenomena such as disequilibria or complexity in molecular distributions. In addition, analogous to the science of *in situ* fossil evidence of life, exoplanet science should similarly account for the preservation, alteration, and destruction ("survivability") of biosignatures at the planetary scale, as well as false negatives and positives, and the imperative that biosignatures are not necessarily single substances but a suite of traits that must be inferred in the context of the stellar-planetary environment.

The OWL decadal strategy was tasked with providing an overview of planetary science, astrobiology, and planetary defense, and ranking the most compelling research questions, challenges, and strategy for 2023-2032. It presented twelve priority science questions based around the formation and evolution of the solar system planets. These questions were applicable to astrobiology and exoplanet science, for which the report gave an abbreviated treatment focused on insights to be derived from Earth as distilled from the previous consensus documents. Most recently, the NASA Exoplanet Exploration Program (ExEP) annual Science Gap List (Stapelfeldt & Mamajek, 2025, 2026) called for the need for a complete inventory of remotely observable exoplanet biosignatures and their false positives, including both gaseous and surface biosignatures, and encouraging novel biosignatures.

A common theme developing through these survey and consensus reports is the need to explore the multi-parameter phase space of potential star-planetary environments along the continuum from planetary formation, prebiotic conditions, chemical evolution, origin of life processes, organisms, and their diversification, so as to interpret any biotic signal in those contexts. All cite the significance of photosynthesis for transforming a planet, but lack precise details on photosynthesis research, in contrast to specifics on, e.g. the geological record, RNA polymerization, and modes of gene transfer and the nature of the last universal common ancestor (LUCA). Both NASA2015 and OWL present tables of extremophile environmental limits, in which both lack tabulations of observed light limits (spectral and intensity) for Earth photosynthesis. OWL says surprisingly little about photosynthesis, other than citing the still unsettled fossil record of when oxygenic photosynthesis first arose. OWL's Question 9.2b (Q9.2b) calls for research vaguely to distinguish "light-dependent ecosystems" from light-independent ones with regard to their potential size in isolation or interactive with each other, but then Q9.2c on fluxes that control productivity leaves out photosynthetically active radiation as one of these controllers.

Exploration of potential photosynthetic biosignatures on other planets thus far has been largely driven by astronomers seeking to benchmark detectability of atmospheric gases (Segura et al., 2005); oxygen as a biosignature (reviewed by Meadows et al. (2018)); surface photopigments on a fiducial Earth-through-time (Kawashima & Rugheimer, 2019) (LUVOIR Final Report, 2019; (Lustig-Yaeger et al., 2023) (Parenteau et al., 2026) or best-case scenario cover of Earth-like organisms on idealized theoretical exoplanets (Borges et al., 2024; Metz et al., 2024). Studies to address biosignature variants from photosynthesis adapted to alternative exoplanet environments have been far and few between. These tend to be of two kinds: studies on the performance of Earth photosynthetic bacteria under the light of other stars (Battistuzzi et al., 2023; Ritchie et al., 2017), or theoretical studies on photopigment optimization to the Sun and other stellar spectra (Arp et al., 2020; Kiang, Siefert, et al., 2007; Lehmer et al., 2021; Marosvölgyi & Gorkom, 2010; Milo, 2009), with the very different stellar spectra of M dwarf stars motivating hypothetical scenarios of NIR oxygenic photosynthesis(Kiang, Segura, et al., 2007; Takizawa et al., 2017; Tinetti et al., 2006), dominant anoxygenic photosynthesis (Coelho et al., 2024; Sanromá et al., 2014), or widespread rhodopsin-based proton pumping (DasSarma & Schwieterman, 2018). Studies that propose alternative configurations at the molecular scale of pigments or photosystems are beginning to emerge (Chitnavis et al., 2024; de la Concepción et al., 2022; Gray et al., 2025; Takizawa et al., 2017). While several efforts have identified planetary false positives and negatives for oxygenic photosynthesis (Meadows et al., 2018), the quantification of the same for surface photopigments has only just begun. Questions about origins of key molecules, like the oxygen evolving complex (OEC) and chlorophyll, and enzymes, like Rubisco, in photosynthesis on Earth today still require answers to constrain uncertainties about their likelihood to arise and proliferate on another planet and whether alternative molecules could take their place. **The research community has not yet achieved a predictive science for photopigment spectral adaptations or broached the potential impact of planetary diversity in prebiotic chemistry, orbital dynamics, magnetic field, gravity, and evolution of surface and atmospheric composition on the emergence, character, and spread of photosynthesis.**

Despite the huge impacts of photosynthesis on Earth and its priority for exoplanet observations, the astrobiology community has not fully tapped the rich body of photosynthesis research that can inform the above unknowns. NASA2015's list of photosynthesis questions are still valid today but very broad and do not elaborate on what the key unknowns are. The HWO2025 proceedings include an SCDD from the Living Worlds Surface Biosignatures Task Group, which provides a review of the conceptual framework to date on Earth-as-an-exoplanet through time scenarios of surface phototrophic spectral signatures for observation simulations studies of their detectability by HWO (Parenteau et al. 2025, hereafter "HWO2025-SBTG"). These scenarios were developed based on expert opinion about relevant reflectance signatures by eon, investigating sensitivity of detectability to different prescriptions of end member spectral features, abiotic background, and their cover over an exoEarth. The great uncertainty in all three of these parameters requires advancing process-based understanding more tightly to quantify their plausibility and probability. In the past decade, discoveries have been made about histories of novel far-red oxygenic photosynthetic bacteria, about how the OEC self-assembles with photo-activation, of accessory pigments attached to rhodopsins, and more that open up possibilities for other planets, and more importantly, help constrain them.

NASA-DARES has devised a framework of 5 "Research" science Focus Areas (FA) similar to the six research areas identified in NASA2015, 2 "Exploration" focus areas on mission development and investment, and 2 "Stewardship" areas on supporting researches and connections to society:

FA1. Protometabolism and Synthesis/Function of Macromolecules in Planetary Environments (Origin of Life )

FA2. Abiotic Organic Production and Chemical Evolution within Planetary Environments

FA3. Co-Evolution of Biospheres, Worlds, and Planetary Systems

FA4. Comparative Planetology to Understand Habitability

FA5. Detecting Signs of Living Environments and Living Worlds

FA6. Astrobiology-Focused Mission Approaches and Technology Development

FA7. Investment in Astrobiology Physical and Digital Architecture

FA8. Workforce and Early Career Support

FA9. Astrobiology in Society

Below, we fill in the details to several of these Focus Areas, with regard to photosynthesis along the continuum of pre-biotic origins, chemical evolution, co-evolution of biosphere and planet, and pathways and limits to alternative biochemistries on other planets informed by the diversity on Earth as well as by theoretical, engineering, and lab experimental studies. This white paper is intended to highlight research areas on phototrophy to encourage tight integration with what the astrobiology community needs to know to interpret future observations for signs of life on other planets.

## 3. Key questions for astrobiology of phototrophy

The fundamental characteristics of phototrophy are the acquisition of light energy for conversion to chemical energy and capture of carbon for synthesis of organic molecules, which may result in distinctive waste products. The detectability of a phototrophic biosphere will be a function of the biosphere size and productivity, which to first order will be constrained by the "photosynthetic photon" (PP) supply, and availability of liquid water, reductant if for autotrophy, carbon substrate, and nutrients. That productivity then will be mitigated by degradation and sinks. The detectability of surface light harvesting pigments will also depend on their spectral absorbance peak(s), which are shaped by numerous functions detailed later. Key broad questions guide this white paper, regarding origins, diversity, process controls, and scaling to the planetary scale through time:

- **How did the critical components of phototrophy on Earth originate?**
- **What is the diversity of phototrophy on Earth?**
- **To what extent do the photosystems observed on Earth reflect fundamental energetic constraints versus historical contingencies of evolution and competition?**
- **Can photosynthesis exhibit biosignatures that are agnostic to the specific chemical composition that produces them?**
- **What affords greater productivity of phototrophy in different environments to increase its detectability to exoplanet observing missions?**

For these broad questions, we have identified specific questions at the cutting edge of understanding to date. In each section below, we cover:

- The importance of the question/phenomenon/method for the search for life, with regard to how it will enhance our ability to constrain the likelihood for it to arise on another planet and/or interpret observations for the presence of biosignatures.
- The key unknowns (and key knowns)
- Recent progress in last 10 years (or major in last 20)
- Types of cross-disciplinary collaborations and/or methodologies/techniques with promise to advance understanding.

## 4. Origins on Earth

There remain important unknowns about the origins of phototrophy on Earth, including the timing of the earliest oxygenic photosynthesis, the relative timing of rhodopsin-based phototrophy, and the origins of key molecules or processes. Such knowledge would provide essential information to be able to conjecture about their possible origins and potential evolutionary paths on other planets. In the late 1970s and early 1980s, well before "astrobiology" was a term for NASA, Bill Schopf organized a consortium of researchers, "Pre-Cambrian Paleobiology Research Group" (Schopf, 2021), which motivated work of Bev Pierson, John Olson, and others, bringing together microbiologists, paleontologists, and others for interdisciplinary examination of the micro-fossil record. Since then, there has been a decades-long gap in pre-biotic origin studies for photosynthesis, while phylogenomic methods to trace back to the last universal common ancestor (LUCA) for photosynthesis have made impressive strides in recent efforts. Below we summarize key uncertainties that are critical for the search for phototrophic life, regarding the early evidence for photosynthesis, the dominant photopigments chlorophyll and rhodopsin, the reaction centers, and RuBisCo.

### 4.1. Geological, geochemical, and phylogenomic evidence for early photosynthesis

Contributors: Greg Fournier, Tim Lyons, Saleheh Ebadirad, Anastasia Yanchilina

**Q4.1a: What geological, geochemical, and phylogenomic evidence do we have now for early photosynthesis, and what still needs to be constrained? (Link to: Section on exoplanet environments)**
**Q4.1b: What delayed the rise of atmospheric $O_2$ after the emergence of oxygenic photosynthesis?**

The origin of photosynthesis remains poorly constrained by the geological and geochemical record (Lyons et al., 2014; Schopf, 2021). Most of these signals are indirect and may be affected by diagenesis, metamorphism, weathering, or later microbial overprinting; consequently, they generally provide minimum-age constraints on the environmental expression of a metabolism rather than directly dating its evolutionary origin (Fischer et al., 2016; Javaux, 2019; Planavsky et al., 2020). The best proxy records consist of the oxidation of potential electron donors, or secondary oxidation of metals by the oxygen produced by oxygenic photosynthesis. The earliest proposed evidence consists of extensive precipitation of banded iron formations during the Archean as old as 3.77 Ga. These have been attributed to anoxygenic phototrophy, supported by stable iron isotope fractionations (Czaja et al., 2013), although abiotic explanations have also been proposed for their formation (Dodd et al., 2022). Archean stromatolites and laminated microbial-mat structures as old as ~3.4 Ga provide complementary geological evidence for early shallow-water microbial ecosystems, some of which have been interpreted as photosynthetic; however, they are not independently diagnostic of phototrophy or of oxygenic versus anoxygenic photosynthesis (Allwood et al., 2006; Tice & Lowe, 2004). The proposed geochemical record for oxygenic photosynthesis also begins in the Archean, around 3 Ga, and consists of oxidation of many different trace metals that would require $O_2$ or a similarly strong oxidant to form (Crowe et al., 2013). Independent phylogenomic evidence also supports an Archean origin of oxygenic photosynthesis. Fossil-calibrated molecular clocks generally show oxygenic Cyanobacteria diversifying at least several hundred million years before the Great Oxidation Event (GOE) (Boden et al., 2021; Fournier et al., 2021; Sánchez-Baracaldo & Cardona, 2020). This suggests that oxygenic photosynthesis predated the atmospheric rise of oxygen by hundreds of million years, indicating that lack of $O_2$ detection as a biosignature may lead to false negatives (Lyons et al., 2014; Reinhard et al., 2017), and that additional planetary processes/factors are required for atmospheric $O_2$ to accumulate.

The delayed rise of Earth's atmospheric $O_2$ likely reflects the balance between biological $O_2$ production and large, evolving oxygen sinks, including reduced volcanic gases, Fe(II)-rich oceans and crust, and atmospheric methane. Ecological competition with anoxygenic phototrophs and methane-cycle feedbacks may have further limited net $O_2$ accumulation before the GOE (Ebadirad et al., 2026; Lyons et al., 2014; Lyons et al., 2024; Ozaki et al., 2019).

Critical uncertainties remain regarding the evolutionary affiliations and metabolic capabilities of the earliest phototrophs, the timing and sequence of the biological innovations leading to oxygenic photosynthesis, whether proposed Archean oxidation signals record localized biological $O_2$ production or later alteration and abiotic processes, and, in general, what delayed the rise of atmospheric oxygen. Resolving these uncertainties will require integrative studies using multiple proxies, and establishment of sound abiotic controls to distinguish primary biological signals from abiotic products and later alteration. This will be aided by development and independent validation of new methods including machine-learning-assisted approaches (Chen et al., 2022; Wong et al., 2025). Furthermore, targeted searches and detailed characterization of exceptionally preserved Archean fossils can potentially more firmly establish an early narrative of photosynthetic life. Improved methods in fossil-calibrated molecular clock analyses can improve the precision and accuracy for determining when different groups of phototrophs emerged, including utilization of genome-derived relative-age constraints such as HGTs, rigorous refinement and testing of alignment datasets, and improved evolutionary models (Fournier et al., 2022). These approaches will be aided by expanded recovery of high-quality genomes and, where feasible, cultivation of deeply branching cyanobacteria, non-photosynthetic cyanobacterial relatives, and poorly sampled phototrophic lineages.

## 4.2. Porphyrins and Chlorophylls

Contributors: Jonathan Lindsey

**Q4.2: How and when did chlorophyll emerge?**

The question of "How and when did chlorophyll emerge?" has been posed over the years without a satisfactory answer. Possible phylogenetic rooting (Cavalier-Smith, 2006) \ of the last common cellular ancestor in the Chlorobacteria does not answer the question, only pushes it deeper in time.

A recurring view since the 1950s is that tetrapyrrole macrocycles were valuable if not essential for the origin of life. If a simple form of photosynthesis (protophotosynthesis) can originate spontaneously and thereby fuel the first cells (protocells), the answer to the aforementioned question lies in the prebiotic era (Soares et al., 2012).

The extant biosynthesis of tetrapyrroles is a paradigm of molecular concision: 2 molecules of delta-aminolevulinic acid (ALA) self-condense to form the pyrrole porphobilinogen (PBG), 4 of which self-condense to form the macrocycle uroporphyrinogen (Urogen), the universal precursor to heme, chlorophylls, bacteriochlorophylls, cobalamin, etc.

Extensive efforts to recapitulate the biosynthesis via structure-directed reactions without enzymes have failed: while PBG spontaneously yields Urogen and therefrom diverse photo/redox-active tetrapyrrole macrocycles, ALA does not similarly form PBG spontaneously (Taniguchi et al., 2017)

Whether photosynthesis came before or after the origin of life is an open questions; regardless, how an early cell acquired the ability to utilize photons is not well constrained. Two ideas put forth are that 1) it was the fortuitous adaptation of primitive pigments involved in infrared thermotaxis of chemolithotrophic bacteria at hydrothermal vents (Nisbet et al., 1995) or 2) it arose first in shallow waters from UV photo-protective pigments that then transferred the energy to the porphyrin (Mulkidjanian & Junge, 1997). Since these two hypotheses, there has been little new to confirm or refute, so the origin of chlorophototrophy remains an open question that identifies a gap between the Origin of Life to the emergence of the last universal common ancestor (LUCA) for photosynthesis.

The study of alternative pathways – compatible with present understanding of planetary geochemistry – to suitable self-condensable pyrroles (Alexy et al., 2015) is of urgent interest, particularly since porphyrins in membranes are capable of diverse interfacial photoreactions (Ilani & Mauzerall, 1981) and proton pumping (Sun & Mauzerall, 1996), both signature photosynthetic features of fundamental prebiotic relevance. Research to identify prebiotic plausible pathways to tetrapyrrole macrocycles would deepen our understanding of the origin of life, the origin of photosynthesis, and perhaps their interplay.

### 4.3. Proton-pumping rhodopsins

Contributors: Massimo Olivucci, Keiichi Inoue, Oded Beja

**Q4.3a: How and when did retinal and rhodopsin emerge?**
**Q4.3b: Could rhodopsin-based proton-pumping have emerged prior to oxygenic photosynthesis?**

Microbial rhodopsins are currently an extremely diverse group of photoreceptive trans-membrane proteins featuring a single Schiff base retinal chromophore bound to diverse opsins and are found in all domains of cellular life and are even encoded in genomes of viruses. They have a wide spectrum of functions, which include light-driven pumps, channels, sensory rhodopsins and enzyme-rhodopsins (Kato & Inoue, 2026; Rozenberg et al., 2021). Proton-pumping rhodopsins (PPR) on Earth are known in only aerobic heterotrophs, but have been found now in all domains of life. PPRs occur mainly in 3 primary classes: bacteriorhodopsins (BR), proteorhodopsins (pR), and xanthorhodopsins (XR) found in halophiles, marine bacterioplankton and terrestrial microbes (Rozenberg et al., 2021) (Brown, 2014) . However, more classes of PPRs continue to be discovered, including cyanobacterial rhodopsin (CyR) (Hasegawa et al., 2020), and even xenorhodopsins (XeR) and schizorodopsins (SzR) (Kato & Inoue, 2026). Retinal typically requires oxygen for its biosynthesis, but capability for rhodopsin-based phototrophy has been demonstrated in hypoxic and anoxic lab conditions (DasSarma et al., 2012). This leaves open the question whether early retinal or retinal analogue synthesis was necessarily preceded by oxygenic photosynthesis or could have emerged earlier by leveraging other limited sources of oxygen.

There are currently no age constraints on the earliest microbial rhodopsins from the fossil record (Sephus et al., 2022). Phylogenetic reconstructions report that the earliest proton-pumping rhodopsins (possibly *Exiguobacterium sibiricum* Rhodopsin, ESR and Marine *Actinobacteria* Clade Rhodopsin, MACR) are more deeply rooted than the branching of Archaea and Bacteria from a shared ancestor (Shalaeva et al., 2015; Tzlil et al., 2025). A reconstruction of primordial type 1 rhodopsin inferred that it was tuned to a green absorbing light niche (Sephus et al., 2022).

It is possible to support the conjecture that retinoids and their analogues emerged, proliferated, and co-evolved as a response to dissipating the energy of the solar spectrum. In fact, in molecules with conjugated double bonds, the photochemical *cis*–*trans* isomerization is an efficient process for light-energy dissipation (Michaelian & Simeonov, 2015). Furthermore, it is likely that conjugated enones could react with amines to form, in a non-aqueous environment, conjugated Schiff bases, namely, rhodopsin chromophores. Thus, initial biopolymers featuring cavities with an internal aminic function could have bound to conjugated enones to form rhodopsin-like precursors which may have eventually developed a proton-pumping function. However, there is presently no proposed pre-biotic synthesis for, for instance, enones or polyenes with more than two conjugated double-bonds (Cleaves II, 2003). It has been proposed that longer conjugated chains can be synthesized by ribozymes and would therefore be part of the RNA-world (Lilley, 2011; Woerly et al., 2014) (Lee et al., 2019). By coupling with ATPase, proton-pumping rhodopsins can constitute the simplest light-to-chemical energy conversion system and could therefore emerge on planets at an earlier stage than chlorophyll-based photosynthetic systems, which require the evolution of reducing power. In such a system, proton-pumping rhodopsins, that use the proton on their retinal protonated Schiff base chromophores as the photochemical substrate, can efficiently utilize light energy for $H^+$ transfer within a compact molecular architecture.

Given their simplicity, early emergence, and ubiquity in all domains of life, the likelihood for rhodopsin-like proton-pumping to arise on other planets seems high. Future advances to constrain the fossil record, the potential origin pathways of rhodopsins, whether it could precede oxygenic photosynthesis, and its potential signal strength will require continued phylogenomic investigations, surveys to discover novel species, and approaches from theoretical and computational chemistry and biology, laboratory simulation of prebiotic conditions for both Earth and other planets, and collaboration with planetary modelers. Its occurrence in only photoheterotrophic organisms raises the question whether it could support autotrophy, to be treated in Section 6.2.

### 4.4. Reaction centers

Contributors: Christopher Gisriel, Greg Fournier, Colin Gates, Robert Blankenship, Kevin Redding

#### Q4.4: How and when did photochemical reactions centers (RCs) emerge?

Photochemical reaction centers (RCs) are the fundamental molecular machines that convert light energy into stable charge separation and ultimately chemical energy. Because all known chlorophyll-based phototrophy depends on RCs, understanding their origins is essential for assessing how readily phototrophy might evolve on habitable planets and whether photosynthesis should be expected to emerge repeatedly in the universe. While all known RCs on Earth descend from a common ancestor, independently evolved phototrophic systems elsewhere need not resemble terrestrial RCs in detail. Instead, identifying the general principles required for biological photochemistry is critical for evaluating alternative evolutionary solutions.

Several key features of RC evolution are reasonably well established (Hohmann-Marriott & Blankenship, 2011). All known photochemical RCs share a conserved architecture and likely descended from an anoxygenic, homodimeric ancestor capable of photochemical charge separation using a chain of redox cofactors (Orf et al., 2018; Sadekar et al., 2006). However, major uncertainties remain regarding its evolutionary origins and complexity. It is unclear whether the ancestral RC more closely resembled modern Type I or Type II RCs, whether it could perform both quinone- and ferredoxin-linked electron transfer, what pigments it employed, and how antenna systems evolved. The relationship between chlorophyll- and bacteriochlorophyll-based phototrophy remains uncertain, as biosynthetic pathways suggest conflicting interpretations regarding which pigment type arose first. A major additional question is how Type I and Type II RCs became coupled to support oxygenic photosynthesis and whether oxygenic photosynthesis necessarily requires two photosystems.

Recent progress has come from comparative genomics, structural biology, phylogenetics, and pigment biochemistry. Future advances will require integrating these approaches with synthetic biology, protein design, geochemistry, and exploration of poorly characterized phototrophic lineages. Identifying the minimal requirements for a functional RC will help determine whether phototrophy is a rare evolutionary innovation or a predictable outcome of life exposed to light.

### 4.5. Water oxidation

Contributors: Christopher Gisriel, Colin Gates, Greg Fournier

#### Q4.5: How and when did water oxidation and the oxygen-evolving complex emerge?

The origin of oxygenic photosynthesis is among the most important unresolved questions in astrobiology because molecular oxygen is one of the strongest and most readily detectable planetary biosignatures. On Earth, atmospheric $O_2$ is produced almost exclusively by Photosystem II (PSII), which oxidizes water using a unique $Mn_4CaO_5$ catalytic cluster. Understanding how this catalyst evolved is therefore critical for constraining the likelihood that oxygen-producing biospheres arise elsewhere.

Several key observations are well established. PSII is the only known biological water-oxidizing catalyst, and oxygenic photosynthesis appears to have evolved only once. However, major uncertainties remain regarding both the timing and evolutionary pathway of water oxidation. One hypothesis proposes that ancestral reaction centers functioned initially as light-driven Mn oxidases before acquiring the ability to oxidize water (Chernev et al., 2020; Lingappa et al., 2019). This idea is supported by the chaperone-free, light-driven assembly of the modern oxygen-evolving complex and by observations that PSII can generate extended manganese oxides, requiring spatiotemporal constraint of metal oxidation (Oliver et al., 2022). Alternative models propose that water oxidation originated relatively early during reaction center evolution, although the large evolutionary distances separating extant reaction centers make reconstruction of ancestral states particularly challenging (Cardona et al., 2019).

Recent progress includes advances in structural biology, geochemical studies of ancient manganese cycling, experimental investigations of manganese oxidation, and mechanistic studies of photoactivation and metallocluster assembly. Future progress will require cross-disciplinary efforts integrating evolutionary biology, geochemistry, structural biology, synthetic biology, and biophysical studies of metallocluster assembly. Determining whether water oxidation evolved through accessible intermediates or represents a rare evolutionary innovation remains central to predicting the prevalence of oxygenic photosynthesis beyond Earth.

### 4.6. Rubisco

Contributors: Beau Dronsella, Helena Schulz-Mirbach, Carolina Gutierrez-Marinez

**Q4.6: How did Rubisco originate?**

Answering this question would inform about potential pathways for the emergence of autotrophy on other planets, any inherent limitations, and potential productivity that would affect detectability.

Ribulose-1,5-bisphosphate carboxylase/oxygenase (Rubisco) is the central enzyme of the Calvin–Benson–Bassham cycle and dominates carbon fixation in Earth’s biosphere (Erb & Zarzycki, 2018; Prywes et al., 2023). The ancestral origin of Rubisco itself remains unresolved (Amritkar et al., 2025). Rubisco-like proteins, which share homology with true Rubisco but often function in other metabolic pathways such as sulfur metabolism or methionine salvage, may preserve clues about the pre-carbon-fixation history of the enzyme family (Ashida et al., 2005; Tabita et al., 2007). Identifying which molecular changes transformed ancestral Rubisco-like proteins into true $CO_2$-fixing Rubiscos is critical for reconstructing the evolutionary pathway from general metabolic enzymes to globally important carbon-fixation machinery.

Rubisco is found across all domains of the Tree of Life, but its evolutionary history remains incompletely resolved. Forms I, II, and III Rubisco share common ancestry but differ substantially in structure, catalytic rate, $CO_2/O_2$ specificity, and ecological distribution (Tabita et al., 2008). These differences likely reflect adaptation to distinct environmental regimes, including variation in $CO_2$ availability, $O_2$ concentration, temperature, redox state, and carbon-concentrating mechanisms (Flamholz et al., 2019; Iñiguez et al., 2020). Yet the evolutionary transitions that produced this functional diversity remain poorly understood. (Kędzior et al., 2022) performed phylogenetic reconstruction of an ancestral I Rubisco and engineered it into a modern cyanobacteria; they found it to have strongly reduced catalytic kinetics and greater carbon isotope fractionation compared to the wild type (WT).

It remains unclear how Rubisco's oxygenase activity arose, whether it reflects an origin under early anoxic or low-$O_2$ conditions or alternative metabolic functions (Bloom & Lancaster, 2018; Busch et al., 2018), and what adaptations first allowed Rubisco-dependent organisms to remain productive as oxygen accumulated in local environments and, later, in the atmosphere. Resolving this problem is essential for understanding how carbon fixation and oxygenic photosynthesis could become linked in the evolution of Earth's biosphere.

Future advances to address Rubisco's origins would entail studies that address its deep evolutionary history, for instance, through the development and the application of computational approaches and homology-based analyses that are less sensitive to evolutionary distance.

## 5. Light harvesting

Q5: Can light harvesting be considered a producer of agnostic biosignatures?

### 5.1. Primacy of chlorophototrophy and potential for alternatives

Contributors: Kevin Redding, Jonathan Lindsey, Yu Komatsu

#### Q5.1: Is chlorophyll ultimately the best molecule for light harvesting and charge separation?

Chlorophyll has been argued to be the best possible molecule for light harvesting and charge separation (Björn et al., 2009; Mauzerall, 1973) but are there other photopigment molecules that could have precedence in other planetary contexts?

The requirements for a photopigment are daunting: spectral coverage, sharp absorption bands, sufficient excited-state lifetime, suitable reduction potentials of radical states, chemical stability, feasible biosynthetic pathways, self-assembly with macromolecules, and tailorability of all such properties. Considerations of plausible pigments (Michaelian & Simeonov, 2015; Rodriguez et al., 2024)(Rodriguez et al., 2024; Michaelian and Simeonov, 2015) must consider the spectral features (Björn, 1976) as well as all of the aforementioned features.

Proton pumping using conjugated polyenes (e.g., retinal) constitutes an alternative to chlorophototrophy. Yet the spectral coverage of polyenes may be limited more to the visible region (Niedzwiedzki et al., 2026), whereas bacteriochlorophylls absorb across the ultraviolet, visible, and near-infrared (700–1100 nm). Tetrapyrroles can engage in redox reactions that give rise to proton pumping as part of a larger photosynthetic system, whereas polyenes engage only (to our knowledge) in proton pumping as part of relatively simple proteins. Utilization of the resulting chemiosmotic potential requires advanced biochemical pathways, but these were likely developed in early-arising chemolithotrophs.

In near-infrared-rich illumination, such as that expected on some planets orbiting M stars, longer-wavelength-absorbing pigments would become particularly relevant. Linear polyenes offer one route through extended π conjugation, whereas cyclic macrocycles provide a compact, metal-binding framework in which absorption and redox properties can be tuned together. Phthalocyanine analogues or expanding porphyrinoids therefore would merit consideration beyond chlorophylls/bacteriochlorophylls (Komatsu & Takizawa, 2021; Osuka & Saito, 2011; Sessler & Seidel, 2003).

Research priorities to identify other photopigments:
1. Investigation of polyene–macromolecule assemblies for proton pumping features would evaluate suitability in prebiotic venues.
2. Quantitative assessment (light-harvesting, growth parameters, etc.) of extant retinal-based organisms would provide the foundation for understanding ecological and perhaps planetary advantages.
3. Vast literature has been accumulated concerning pigments, yet holistic evaluations are challenging. Development of AI-based systems to search the literature for spectra and physical parameters would afford databases that enable evaluation of diverse pigments.

### 5.2. Light-harvesting pigments

Contributors, Photosynthesis: Christopher Gisriel, Ligia F. Coelho, Roberta Croce, Nicoletta La Rocca, Colin Gates, Elisabetta Liistro, Alessandro Agostini, Daniel Canniffe, Hazel Barton, Nancy Kiang
Contributors, Rhodopsin: Massimo Olivucci, Keiichi Inoue, Oded Beja

The phototrophic photon (PP) supply is characterized by both spectral quality, what photons are useful, and PP flux density (PPFD) quantity, the minimum that can support either successful electron transfers or the compensation point for growth versus the maximum or saturating limit with photodamage. Phototrophs are both adapted to as well as can acclimate to these limits. The detectability of surface light harvesting pigments, as noted earlier, will also depend on their spectral features, where the absorbance peaks can be due to multiple types of pigments serving different functions in a phototroph:

- The primary photopigment(s) responsible for charge separation.
- Antenna pigments that may comprise an array of accessory pigments that fill in the absorbance spectrum and sometimes extend it longer than the primary photopigment through thermally activated uphill energy transfer.
- Photoprotective pigments that serve as anti-oxidants or sunscreens.

This white paper focuses on the first two of these, while for photoprotective pigments the reader can refer to the the review of Schwieterman et al. (2015) and studies of their potential detectability (e.g. Rastogi et al., 2014, doi: 10.1016/j.jphotobiol.2014.09.020). The primary photopigment sets the peak efficiency for the photon conversion to useful energy storage across the radiation spectrum.

### 5.2.1. Spectral peaks and bands

**Q5.2.1a: What is the long wavelength limit for oxygenic photosynthesis or photosynthesis at all?**

**Q5.2.1b: What are the spectral limits to rhodopsin-based proton pumping?**

**Q5.2.1c: Can we formulate a predictive science for the wavelength of peak absorbance of a phototroph as a whole organism in arbitrary environments?**

The wavelengths capable of supporting phototrophy arise from the interplay between universal physical constraints and evolutionary adaptation. At the molecular level, the energetics of charge separation, energy transfer, and electron transfer define the minimum energy required for photochemistry, while the spectral distribution and intensity of available light determine the selective pressures acting on light-harvesting systems. Although the physical principles governing photochemistry are universal (Shockley & Queisser, 1961), biological solutions exhibit remarkable diversity through evolutionary tuning of pigments, pigment-protein interactions, and antenna architectures. A central question for astrobiology is therefore whether phototrophic systems can continuously adapt to virtually any spectral environment, or whether fundamental biochemical and energetic constraints create discrete spectral regimes that define the limits of viable phototrophy. Below, we summarize recent advances in understanding the spectral diversity and limits of Earth's known phototrophic metabolisms.

**Oxygenic photosynthesis (OP)** faces a particularly stringent challenge for useful photon wavelengths, because water oxidation requires exceptionally high redox potentials (Lubitz et al., 2019). In oxygenic photosynthesis, it was long held that water oxidation required chlorophyll *a* (Chl *a*) absorbing at 680 nm in Photosystem II (PSII) to excite an electron from a highly oxidizing ground state potential for water oxidation to a reducing potential for electron transfer reactions, while PS I required the energy of 700 nm (Björn et al., 2009; Mauzerall, 1973). Thus, the spectral bounds of OP have historically been considered as 400-700 nm for “photosynthetically active radiation” (PAR), with the lower bound due to damaging effects of UV-A. However, at the upper bound, in eukaryotes, some algae are able to tune antenna Chl *a*’s connected to both PS I and PS II to as long as ~740 nm. In plants, red forms of Chl *a* excitonically coupled in PS I transfer energy from as long as ~750 nm (Toporik et al., 2020; Wolf et al., 2018), such that suggestions are emerging to expand PAR for plants to 400-750 nm (Amelii et al., 2026; Zhen & Bugbee, 2020).

The uniqueness of Chl *a* was overturned by the discovery of far-red chlorophylls, Chl *d* in *Acaryochloris* (Miyashita et al., 1997; Miyashita et al., 1996), where it replaces almost all Chl *a* in PSII, serves as the electron donor (Renger & Schlodder, 2008), and shifts the PSII peak absorbance to as long as 723 nm (Mielke et al., 2013) and PS I to 740 nm (Hu et al., 1998), with the OP long wavelength "red limit" up to ~760 nm (Oliver et al., 2026). The discovery of Chl *f* (Chen et al., 2010) has further extended the possible long wavelength limits for OP. In Far-Red Light Photoacclimation (FaRLiP) (Gan et al., 2014; Ho et al., 2017), Chl *d* and Chl *f* are minor pigments induced in response to far-red light enrichment, extending PSII to ~727 nm and the PAR upper range to ~800 nm (Nürnberg et al., 2018).

**Anoxygenic photosynthesis** uses reduced organic or inorganic compounds such as sulfide, ferrous iron, or hydrogen as electron donors, instead of water, not producing oxygen. These systems use bacteriochlorophylls to absorb in the near infrared, with reaction center trap wavelengths ranging 800-960 nm (for a table of photopigments, trap wavelengths, and electron donors, see Kiang (2014)). The current upper spectral bounds come from purple non-sulfur bacteria, specifically *Blastochloris viridis* (formerly *Rhodopseudomonas viridis*, Hiraishi (1997)), with maximum absorbance wavelength measured at 1018-1020 nm (Drews & Giesbrecht, 1966; Namoon et al., 2022). But this limit is empirical and absorption further in the IR remains a research question to pursue (Kimura et al., 2022).

**Rhodopsin-based proton-pumping.** Rhodopsins can work in three different "molecular regimes": i) with a protonated chromophore (Karasuyama et al., 2018; Sacchetta et al., 2025), ii) with an unprotonated chromophore (Galindo et al., 2025; Lamm et al., 2025), and iii) using energy transfer from a carotenoid molecule (Balashov et al., 2005; Chazan et al., 2023; Fujiwara et al., 2025; Tzlil et al., 2025). This gives rhodopsin systems a wide flexibility in terms of light harvesting as well as (in case iii) of absorption band broadening. While the peak absorption wavelengths of known proton-pumping rhodopsin range 488–570 nm (Hasegawa-Takano et al., 2024; Nagata & Inoue, 2021; A. Pushkarev et al., 2018), the retinal chromophores exhibit, in its unprotonated or protonated forms, absorption at 370–720 nm when tuned by the opsin cavity (Broser et al., 2023; Galindo et al., 2025), and it is not impossible in principle that this contributes to $H^+$ transfer distinct from that found in known proton pumps.

**Accessory pigments** reshape spectral and flux limits, as well as buffer against oxidative stress, including carotenoids (across all phototrophs, 400 to 600 nm), phycobilins (in cyanobacteria and some algae, 500 to 600 nm), and Chl *b* (plants and algae), Chl *c* (algae), and Chl *d* and Chl *f* (FaRLiP cyanobacteria). These pigments in some organisms can always be present, in others only expressed in certain conditions, the trade-offs for which bear further quantification.

#### 5.2.2. Flux limits

**Q5.2.2: What are the low and high light flux limits for different kinds of phototrophy?**

In photosynthesis, the currently known most extreme example of growth in low light flux is green sulfur bacteria. These anoxygenic phototrophs contain giant light-harvesting antenna assemblies called chlorosomes, allowing them to survive in minimal light, such as that from black-body radiation from hydrothermal vents rather than light from our Sun (Beatty et al., 2005). On the other end of the spectrum, some cyanobacteria have adapted to extremely high light fluxes, especially those isolated from deserts (Dobson et al., 2021; Levin et al., 2021). Understanding the molecular mechanisms by which photosynthetic organisms tune their light usage is necessary to determine the minimum and maximum photon flux under which photosynthetic life, and by extension consumers, can be expected to develop and survive.

In a system reliant on one-electron processes, as with all terrestrial reaction centers other than PSII, the minimum photon flux is dependent on the efficiency of charge separation and the maintenance energy cost of the organism. The minimum photon flux for PSII or any theoretical reaction center with a cofactor capable of accumulating multiple charges is constrained by the recombination/neutralization rate, which is not well understood. The maximum survivable photon flux for all reaction centers is determined by factors including 1) the consequences of failure to charge-separate; 2) the availability of electron sources and sinks (diffusion and quantity of donors/acceptors, ability to produce or acquire antioxidants); and 3) the presence of photoprotective mechanisms including non-photochemical quenching (NPQ) (Zuo, 2025). For terrestrial oxygenic photosynthesis, the minimum observed survivable light flux is 0.04 mmol photons $m^{-2}$ $s^{-1}$ (Hoppe et al., 2024), while the maximum is less clear, as plants and algae have been lab-adapted to survive at intensities above native sunlight (Treves et al., 2016).

Current understanding of minimum and maximum light flux for photosynthesis has been shaped by field and laboratory biology, as well as computational modeling. These methods must be continued and supplemented with omics studies and structural, synthetic, molecular, and cell biology approaches. Understanding the differences between theoretical and practical limitations will constrain which stars and planets should be targeted in the search for life.

**Rhodopsin-based proton pumping** excels at increasing the production rate of ATP under high light, due to the ability to accelerate the photocycle turnover rate, the mechanisms for which are well understood (REF). The extreme light limits for rhodopsin-based proton pumping has not been investigated, but in marine algae light saturation was observed to occur not until 2000 μmol/m2/s in 488-570 nm (Andrew et al., 2023), where the maximum total visible flux density on Earth under clear skies at the equator is 2000-2200 μmol/$m^2$/s. In terms of low light adaptation, the limit imposed by rhodopsin quantum efficiency can be up to 75% above vs. thermal noise (dark state isomerization)(Gozem et al., 2012), but to our knowledge the low light flux limit has not otherwise been investigated.

### 5.2.3. Adaptation and acclimation

**Q5.2.3a: What are the trade-offs of adaptation or acclimation to alternative photon wavelengths?**

**Q5.2.3b: What are the trade-offs of adaptation or acclimation to low/high light?**

**Q5.2.3c: Can light harvesting pigment spectral tuning to the spectral light environment be considered an agnostic biosignature?**
**Q5.2.3d: What light harvesting capability would be competitive (most productive) in alternative environments?**

Organisms are evolutionarily adapted to spectral and flux light niches and, on short time scales, acclimate to fluctuating light conditions. Adaptation alters pigment composition, pigment-protein spectral tuning, antenna architecture, photosystem stoichiometry, electron transport pathways, and photoprotective mechanisms over evolutionary timescales. In rhodopsins it alters not only the opsin sequence but also the chromophore itself that becomes deprotonated to adapt to shorter wavelengths and increases the conjugated chain lengths (A2 chromophores) to adapt to longer wavelengths (Luk et al., 2016). Acclimation allows organisms to respond to changing environments on timescales ranging from minutes to days, commonly with far-red adaptations and/or production of accessory pigments and, in plants, alteration of carboxylation capacity. At low irradiance or low photon energies, productivity is constrained by the efficiency of charge separation, maintenance-energy requirements, and the stability of photochemical intermediates. At high irradiance, productivity is limited by photodamage, reactive oxygen species formation, electron acceptor availability, and the energetic costs of protective mechanisms such as non-photochemical quenching (Zuo 2025).

For oxygenic photosynthesis, recent progress to quantify the trade-offs in adaptation to different light niches, and the trade-offs between adaptation and maintaining acclimation machinery has been through the discoveries of the diversity of far-red phototrophs (Antonaru et al., 2020), characterization of spectral diversity within chlorophyll *d*-utilizing cyanobacteria (Elias et al., 2024), and increasing integration of structural, physiological, and ecological studies of photosynthetic adaptation (Ulrich et al., 2024; White et al., 2025). While a growing number of far-red oxygenic phototrophs are being cultivated in the laboratory, more studies are needed that characterize the native light environment, both the spectral quality and photosynthetic photon flux density (PPFD), of phototrophic organisms that inhabit different spectral environments.

However, major uncertainties remain regarding the limits of these responses and the tradeoffs associated with maintaining acclimation machinery (Elias et al., 2024; Viola et al., 2022). Systems such as far-red light photoacclimation require extensive genetic regulation, pigment remodeling, and replacement of photosystem components, raising questions about when such flexibility is advantageous (Gan et al., 2014). Protective responses such as NPQ typically lower momentary photosynthetic efficiency in exchange for reduced damage and sustained productivity over longer intervals, but quantifying the losses to NPQ is not yet an exact science.

### 5.2.4. Key priorities for astrobiology research on light harvesting pigments

This Section 5 on light harvesting has highlighted that interpreting remotely detectable phototrophic biosignatures requires understanding not only the fundamental mechanisms of biological light harvesting, but also the evolutionary and environmental processes that shape them. Recent discoveries have substantially expanded the known diversity of phototrophic strategies, including far-red oxygenic photosynthesis, novel rhodopsins, and increasingly diverse pigment systems, demonstrating that terrestrial phototrophy occupies a broader range of spectral and ecological niches than previously appreciated. At the same time, fundamental questions remain regarding the physical limits of phototrophy, the evolutionary pathways by which these systems arise, and the extent to which photosynthetic traits can be predicted from planetary environments.

Advances that will substantially improve our ability to assess the likelihood of photosynthetic life on other worlds and to interpret future observations of potentially inhabited planets include  the priority research areas:

- Establishing the true spectral and flux limits of oxygenic and anoxygenic photosynthesis and proton-pumping rhodopsins through complementary laboratory, field, and theoretical; structural, physiological, and computational studies.  These efforts should rigorously quantify both spectral quality and photon flux under environmentally relevant conditions.
- Discovering and theorizing novel forms of phototrophy at spectral and flux extremes or coupled with alternative carbon fixing pathways (see Section 6.2).
- Quantifying how atmospheric and surface environmental conditions shape spectral adaptation and costs of spectral acclimation.
- Determining how atmospheric composition, stellar spectra, and surface physico-chemical environments shape photosynthetic adaptation and acclimation, including the energetic costs of maintaining flexible light-harvesting strategies
- Demonstrating the viability of spectral tuning via multi-photosystem electron transfer pathways (see Section 5.3).
- Integrating molecular biology, evolutionary biology, ecology, spectroscopy, structural biology, planetary science, and predictive modeling will be essential for linking environmental conditions to phototrophic evolution, biosphere productivity, and remotely detectable biosignatures.

### 5.3. Optimization of the electron transfer pathway

Contributors:  Nathan Ennist, Gary F. Moore, Colin Gates, Felisa Wolfe-Simon, Kenji Takizawa, Junko Yano, Kevin Redding

**Q5.3a: How is coupling of the electron transfer chain (ETC) and the generation of the proton gradient for ATP production affected by differences in both light quality and quantity?**

**Q5.3a-1: What tradeoffs exist among energetic efficiency, metabolic flux, photoprotection, spectral coverage, photo-excited state energies of pigments, and the number of photosystems in an electron-transfer pathway?**

**Q5.3b: What electron-transfer architectures beyond those found on Earth are physically and biologically feasible?**

**Q5.3b-1: Can photosynthesis occur without membranes in a soluble system?**
**Q5.3b-2: Can photosynthesis occur without energy storage that depends on generating proton gradients?**
**Q3.2.3b-3: What alternative electron carriers are plausible?**

**Q5.3c: How does planetary chemistry (i.e. trace metal availability, geochemistry, or availability of electron donors and acceptors) influence the evolution of photosynthetic electron transfer?**

Predicting how photosynthetic reaction centers (RCs) are optimized under different stellar spectra, photon fluxes, and in the presence of different electron donors is essential for assessing the likelihood of extraterrestrial phototrophy. Under low light, natural selection may favor photosystems with high thermodynamic efficiency and broad spectral absorption, while high-light conditions may favor photosynthetic pathways that support high metabolic flux without sacrificing energetic efficiency (Burnetti et al., 2026).

Electron transfer pathways on Earth involve membranes that allow proton pumping against a chemical gradient to store energy. Membrane localization also enables the use of hydrophobic quinones as mobile electron carriers, reducing charge recombination by restricting diffusion of electron carriers to the membrane region. Furthermore, the evolution of internal membranes (e.g., chromatophores, thylakoids) allowed greater control of proton gradient formation (i.e., pumping into a small volume rather than the external environment) and the redox chemistry linked to it.

Whether membrane-free alternatives could function effectively remains an open question, one that may be addressed through computational modeling and synthetic biology approaches. One may envision such reaction centers linked to enzymatic activities at each end, one using the holes and the other using the electrons created by charge separation; i.e., being used to drive difficult redox chemistry without contributing to proton gradient formation. Moreover, such a design might be compatible with use of different pigments in distinct reaction centers driving different redox chemistries. Testing the feasibility of membrane-free photosynthetic mechanism via synthetic biology and simulations could expand potential pathways for the early emergence of photosynthesis.

Alternatives to Earth's two-photosystem Z-scheme in oxygenic photosynthesis have been proposed for better optimization to different stars and may include (1) altering cofactors and their redox potentials as observed in far-red cyanobacteria (Viola et al., 2022) and *Gloeobacter* (Kato et al., 2022), or in bioengineering of artificial systems (Ennist et al., 2022); (2) spectral separation of PSI and PSII (Blankenship et al., 2011; Takizawa et al., 2017); and (3) 3-photosystem schemes (Wolstencroft & Raven, 2002). Multi-photosystem schemes may be suited to environments with more abundant low-energy photons, but viable quantitative efficacy has yet to be demonstrated.

Key priorities: Identifying evolutionary paths toward such alternative systems and their efficiency tradeoffs is needed to constrain quantitatively what could be plausible in different environmental contexts. Future progress requires integrating recent developments in theoretical modeling of electron-transfer energetics and kinetics with synthetic biology, rapid evolution techniques, and laboratory (re)construction of alternative photosynthetic pathways to determine which RC architectures are physically and biologically feasible. Collaborations among biophysicists, evolutionary biologists, photophysical chemists, and geochemists to identify plausible evolutionary trajectories and reconstruct the emergence of RCs on Earth will be crucial.

### 5.4. Mitigating photodamage

**Q:5.4 How do compensation strategies for dealing with liabilities of light harvesting affect efficiency, e.g. non-photochemical quenching, coping with oxygen?**

Adapting phototrophy involves a fundamental tradeoff: tuning redox potentials for efficient energy capture versus mitigating severe photodynamic damage (Blankenship et al., 2011; Cardona et al., 2012)(Cardona et al., 2012; Blankenship et al., 2011). Complex ET pathways utilize transition-metal cofactors whose extreme potentials inevitably generate destructive radicals, necessitating the parallel (strict?) co-evolution of photoprotective pigments and antioxidant enzymes (Cardona et al., 2019; Foyer, 2018). Conversely, simpler architectures like retinal-based rhodopsins bypass ET-induced radical stress entirely, trading broad redox capacity for robust energy transduction(DasSarma & Schwieterman, 2021). (Ultimately, it could be that planetary trace-metal availability constrains whether life sustains complex ET networks or defaults to simpler pigment driven pumps. In essence, local trace-metal inventories may well constrain ET cofactor selection and spacing. These metals are critical to charge retention with low size, energetic cost, and reactivity as compared to organic redox mediators.

In the event that ET networks are sustainable, the fundamental efficiency of ET processes will be constrained by the rate of electron sequestration and the capacity for energy dissipation. While many photoprotective mechanisms exist in oxygenic photosynthesis, including NPQ, cyclic electron flow, water-water cycling, photorespiration, and simply replacing damaged components, the molecular mechanisms for almost all of these are not completely understood (Pinnola & Bassi, 2018). Many of these processes have been discovered via spectroscopy and protein knockouts.

Key priorities are to understand the range of possibility for photoprotective mechanisms of this sort, both extant and theoretical, further study is needed in field and laboratory biology, structural and synthetic biochemistry, bioinorganic chemistry, and computational modeling.

## 6. Carbon fixation limits

Carbon fixation, the reduction of inorganic carbon, $CO_2$, to organic, defines autotrophy, and is central to sustaining biochemical cycling on a planet and the proliferation of life.

This ability to fix $CO_2$ and its controls on the potential productivity of a planet co-evolves with the availability of $CO_2$ and other reactive species with changing atmospheric composition. In the search for life, this potential productivity and its sinks are a strong control on the detectability of a biosphere by exoplanet missions. Below we summarize key efficiency issues of the carbon fixing pathways, and their implications for constraining the productivity and detectability of a biosphere over geological time.

### 6.1. Carbon fixing pathways

Contributors: Beau Dronsella, Helena Schulz-Mirbach, Carolina Martinez-Gutierrez

**Q6.1a: Why does the Rubisco-dependent Calvin-Benson-Bassham cycle dominate on Earth as a carbon fixation pathway?**
**Q6.1b: What could be the theoretical ultimate potential efficiency of carbon fixation on Earth in the future?**
**Q6.1cc: How may carbon fixation efficiency evolve as a planet's CO2 level is drawn down over geologic time with the carbonate-silicate cycle evolution with stellar evolution?**
**Q6.1d: What alternative carbon fixation schemes could be successful and become prevalent on another planet?**

There are seven known metabolic solutions for carbon fixation on Earth, of which only three are known to have been incorporated into photosynthesis (Figueroa et al., 2018; Santos Correa et al., 2023). The Calvin-Benson-Bassham (CBB) cycle with Rubisco by far dominates, accounting for >90% of $CO_2$ fixation into biomass on Earth (Erb et al. 2018). Rubisco is well-known to have the liability of being both a carboxylase and an oxygenase. In addition to fixing $CO_2$, it can react with $O_2$, initiating photorespiration and costing in carbon-fixation efficiency (Flamholz et al., 2019). This dual reactivity creates a central evolutionary paradox: oxygenic photosynthesis, the metabolism responsible for producing most of Earth's $O_2$, generates the same molecule that interferes with Rubisco-mediated carbon fixation. Despite Rubisco's shortcomings, it remains the dominant carbon fixing enzyme on Earth (Löwe & Kremling, 2021). While other carbon fixation routes (e.g. the Wood-Ljungdahl pathway, the reductive TCA cycle, and the dicarboxylate/4-HB cycle) in part employ faster carbon fixing enzymes, they are less $O_2$-tolerant, and are limited to organisms in largely anoxic environments.

To mitigate the oxygen inefficiency, organisms have evolved varieties of Rubisco with trade-offs in $CO_2$ specificity vs. catalytic activity (Flamholz et al., 2019; Iñiguez et al., 2020); architectures for carbon concentrating mechanisms (CCMs); and intermediates like PEP carboxylase for spatial and temporal separation of $CO_2$ capture from its fixation by Rubisco through C4 and CAM photosynthesis. Efforts to overcome this include the reconstruction of ancestral Rubisco variants (Schulz et al., 2022). While much effort h has gone into engineering a better RubisCO, one can consider carbon fixation and energy generation as separate modules with diverse solutions that can be interchanged. Efforts through synthetic biology to replace Rubisco or devise synthetic $CO_2$ fixation cycles are ongoing (Bar-Even et al., 2010) (Schwander et al., 2016) . There have been recent successes at replacing the CBB with the more energy-efficient reductive glycine pathway for growth on formate and $CO_2$ in a facultative heterotroph/autotroph, the hydrogen-oxidizing *Cupriavidus necator* (Dronsella et al., 2025); and redirecting the photorespiration pathway (Lu et al., 2025; South et al., 2019). The efficacy of substituting these engineered solutions is limited by the highly integrated nature of biological systems. Will the lifetime of phototrophy on Earth be extended through a more highly efficient engineered or evolved C fixation pathway? What pathway would arise naturally on an exoplanet? Availability of key metals may also play a role in what pathway emerges and dominates. For Earth C fixation pathways these are: Mg (required for enediolate formation in Rubisco), Mn and Zn (carboxylase cofactors), iron (ferredoxin-dependent chemistry), Ni (part of Ni-Fe-S clusters, e.g. in CODH), Co (part of the cofactor $B_{12}$) and Mb (cofactor of some CODH variants).

Key priorities are to theoretically identify possible the carbon fixation pathway solution space for another planet, which will require:

- Integration of geochemistry, biochemistry, molecular biology, computational chemistry and biology, and evolutionary biology.
- Studies to determine what the simplest possible Rubisco would look like and how different Rubisco forms perform under a range of planetary conditions.
- Cross-referencing thermodynamic constraints (atmospheric composition, climate, radiation) with metabolic pathway architectures
- Emerging methodologies, including lab simulation of fast evolution through machine learning to verify alternative feasible solutions.
- Quantifying the availability of metals and metal-like elements would help understand which cofactors for accepting electrons and thus which chemistries are available in the respective environment.

### 6.2. The potential for rhodopsin-based autotrophy

Contributors: Massimo Olivucci, Keiichi Inoue, Oded Beja, Nancy Kiang

**Q6.2a: What limits rhodopsin-based phototrophy from supporting fixation of $CO_2$?**
**Q6.2b: Can rhodopsin-based phototrophy be used to support any of the known carbon fixation pathways or a theoretical construct?**
**Q6.2c: Will rhodopsin-based phototrophy inherently be low productivity and therefore low detectability?**

Proton-pumping microbial rhodopsins support photoheterotrophic archaea and eubacteria, and, although they also occur in eukaryotic algae and cyanobacteria to support ATP production, they are not coupled to carbon fixation pathways. Why they do not directly support carbon fixation for autotrophy remains an open question. The lack of autotrophy possibly limits the large-scale productivity and density of organisms whose sole light harvesting system for energy is rhodopsin. Remote detection of these phototrophs on Earth has been limited to small areas with restricted environmental conditions, such as haloarchaea in salt ponds (Dalton et al., 2009).

However, an artificial photosynthesis system which combines rhodopsin proton-pumping with an extracellular electron uptake mechanism has been reported. This system establishes a pathway to drive photoelectrosynthetic $CO_2$ fixation by *Ralstonia eutropha*, a facultatively chemolithoautotrophic soil bacterium engineered to heterologously express an extracellular electron transfer pathway of *Shewanella oneidensis* MR-1 and *Gloeobacter* rhodopsin (Tu & Huang, 2023). Although rhodopsins are absent from higher plants, proton-pumping rhodopsins are widely distributed among diverse unicellular algae and also occur in cyanobacteria. The expression of some algal proton-pumping rhodopsins is enhanced in iron-deficient environments, where chlorophyll-based photosynthesis is limited (Andrew et al., 2023; Strauss et al., 2023), thereby compensating for reduced ATP synthesis. Rhodopsin-generated proton-motive-force and local acidification promote iron uptake and increase the $CO_2$ concentration in the plastid intermembrane space, respectively, thereby facilitating chlorophyll-based photosynthesis (Strauss et al., 2023; Yoshizawa et al., 2023). It has been reported that microbial rhodopsin powered anaplerotic $CO_2$ fixation could occur in marine bacteria (Pinhassi et al., 2016). However, as initially stated above, despite microbial rhodopsins representing the most widespread phototrophic system worldwide at the genetic level (Finkel et al., 2013), the natural relationship between rhodopsin and microbial autotrophic carbon fixation is still unclear.

Given the simplicity of rhodopsins and how common they are throughout all domains of life on Earth, we cannot ignore the possibility for them or something similar to arise on another planet. To clarify the plausibility of biosphere scenarios in which rhodopsin-like biosignatures might dominate as interpretations of telescope observations, further investigation of novel coupling of rhodopsin-like proton pumping to inorganic carbon fixation together with their environmental constraints should be pursued through theoretical modeling, especially prediction of electronic and vibrational spectra,and bioengineering studies, and continued discovery of novel rhodopsin-based phototrophs.

### 6.3. Continuous photosynthetic habitability

Contributors: RJ Graham, Tim Lyons

**Q6.3a: Can we constrain the continuous photosynthetic habitability lifetime and orbital distance of a fiducial Earth around another star?**

**Q6.3b: Can we constrain the co-evolutionary stage of a planet and biosphere from limited observational data?**

The first question is needed for devising search strategies for habitable and inhabited exoplanets for upcoming telescope missions. The second question is important for interpreting observations for the presence of life according to understanding of self-consistent biosignatures with the planet's evolutionary context.

Carbon fixation on Earth has co-evolved with the atmosphere (Iñiguez et al., 2020; Martin et al., 2018), from early emergence during an Archean anoxic, high-$CO_2$, possibly high-$CH_4$ atmosphere (Catling & Zahnle, 2020), to our modern 20% oxygenated atmosphere (HWO2025), and will continue to do so in the future. The death of Earth's biosphere is thought to be driven by the gradually brightening Sun, with resultant warming and $CO_2$ drawdown by the carbonate-silicate cycle in response (Caldeira & Kasting, 1992; Walker et al., 1981)(Walker et al. 1981; Caldeira & Kasting 1992). Early estimates of remaining photosynthetic habitability gave ~100 Myr (Lovelock & Whitfield, 1982), revised to ~1 Gyr by (Caldeira & Kasting, 1992), who distinguished plant $CO_2$ starvation (~10 ppm) from thermal limits (~323 K). Subsequent 0-D and 1-D studies clustered near 0.5–1.3 Gyr (Franck et al., 2000; Lenton & von Bloh, 2001; Mello & Friaça, 2020; Rushby et al., 2018). Some models permit survival to ~1.6–1.9 Gyr due to weaker weathering feedback (Graham et al. 2024), more robust land plant tolerances (1-3 ppm $CO_2$, 338 K), and more gradual warming in 3D models (Haqq-Misra & Wolf, 2026), which may allow remotely detectable biosignatures to persist longer (Ozaki & Reinhard, 2021).

These estimates to date have progressed from approximate analytical solutions, to 0-D and 1-D geochemical simulation studies, to the moderating effects of 3D climate models simulating time-slice equilibria. The treatment of photosynthetic $CO_2$ fixation in these studies has been limited to $CO_2$ compensation points of reference/model Earth organisms that utilize the Rubisco-dependent Calvin-Benson-Bassham carbon fixation pathway, but with accounting for how Rubisco itself has evolved in response to changing atmospheric CO2/O2 (Iñiguez et al., 2020). Future advances better to constrain the lifetime of a biosphere should take into account the interference of $O_2$ with Rubisco carboxylation, account for theoretical specificity and kinetic limits of alternative carbon fixation pathways as described in Section 6.1 on Carbon Fixation, and explore further evolutionary scenarios with climate-chemistry simulation models of the Earth as well as exoplanets.

## 7. Co-evolution of photosynthesis and environment across time and space

The earlier sections of this paper identified key unknowns about the origin of critical molecules or functions in phototrophy on Earth, and summarized explorations of their potential improvements or replacement with alternative systems given parameters of known mechanisms to date. Here we expand on how these unknowns about origins and mechanisms in the context of other planetary environments scale through evolution and to planetary-scale signatures.

### 7.1. Pathways of evolutionary precedence and ecological dominance

Contributors: Anthony Burnetti, Greg Fournier, Taro Matsuo, Saleheh Ebadirad

**Q7.1a: On Earth, how have the dominant signatures of photosynthesis over the course of co-evolution with the planet relied on selectivity for early biomolecules and earlier successional types of organisms?**

**Q7.1b: On another planet, what other paths could lead to the same outcome or diverge to other dominant signatures at different stages of a planet's evolution?**

**Q7.1c: Does photosynthesis inevitably evolve to become more oxidizing?**

**Q7.1d: How likely should it be for a planet to be dominated by a few photopigments versus exhibit diverse photopigments with more complex biosignatures?**

A planet's dominant phototrophic biosignatures at a point in time need not be the same as the earliest phototrophic innovation or reflect the full diversity of phototrophic pathways present on a planet. Their emergence comprises three stages: biochemical origin, ecological expansion through evolutionary lineage diversification (often mediated by horizontal gene transfer), and eventual expression at planetary scales under particular light, redox, nutrient, and climatic conditions (Fournier et al., 2021; Lyons et al., 2024; Ward & Shih, 2021). Phototrophic systems established early may generate priority effects by occupying bioenergetic and spectral niches, as may be indicated by the ecological complementarity between chlorophototrophy and rhodopsin-based proton-pumping (Burnetti et al., 2026), or the dominance by Rubisco despite its inefficiencies upon the rise of atmospheric oxygen (Amritkar et al., 2025). Subsequent innovations could further reshape the selective landscape, as oxygenic photosynthesis did on Earth by altering electron-donor availability, nutrient cycling, atmospheric composition, and aquatic light environments (Kiang, Segura, et al., 2007; Lyons et al., 2024; Ozaki et al., 2019; Ward & Shih, 2021). These feedbacks may suppress pre-existing metabolisms while creating opportunities for others, and environmental disturbances may periodically weaken dominant lineages and allow previously rare pathways to expand. As such, lineage diversification and horizontal transfer may increase biochemical and taxonomic diversity over evolutionary time, whereas ecological competition and environmental filtering may limit the number of phototrophic strategies that become dominant; e.g. modern anoxygenic phototrophs may have been pushed to marginal environments from an initially more widespread distribution(Burnetti et al., 2026; Ozaki et al., 2019; Ward & Shih, 2021). Consequently, older planets may not display progressively more complex photopigment biosignatures, but instead signatures of phototrophs that marginalize others. Even where underlying pigment diversity is high, limited surface coverage, atmospheric transmission, clouds, and observational resolution may restrict remote detectability to only a few dominant signals. Different sequences of biological innovation and environmental change could therefore produce similar atmospheric biosignatures but distinct surface-reflectance signatures, or conversely, similar surface signatures but different atmospheric outcomes.

Testing these ideas will require an integrative framework, starting with formulating origin scenarios, including pre-biotic chemical evolution to characterize the probabilities for emergence of specific biological molecules as well as comparative phylogenomic analyses that reconstruct backward in time the distinct evolutionary histories of reaction centers, pigment-biosynthesis pathways, electron-donor utilization, carbon-fixation systems, and photoprotective traits, while molecular-clock analyses constrain the timing of their origins and diversification (Fournier et al., 2021; Ward & Shih, 2021). Ecological and biogeochemical models can then identify the environmental conditions under which these innovations could spread and become dominant (Ozaki et al., 2019), whereas spectral biosignature models can translate their abundance and spatial distribution into remotely observable signals (Hegde et al., 2015; Kiang, Segura, et al., 2007; Schwieterman et al., 2015). This approach could distinguish the initial origins of phototrophic innovations from their subsequent horizontal transfer and ecological expansion, and identify the conditions under which they became sufficiently widespread to generate planetary-scale biosignatures. Experimental studies of spectral performance and competitive trade-offs across gradients in light quality, redox state, electron-donor availability, and nutrient supply could further test whether historical contingency produces persistent ecological dominance or whether similar environmental constraints repeatedly favor convergent phototrophic strategies (Burnetti et al., 2026; James et al., 2006).

Key priorities for research to constrain the path-dependent co-evolution of photosynthesis with the planetary environment should engage in cross-disciplinary collaborations and methodologies that develop integrative frameworks to consider one or more of the these five interacting processes:

- Pre-biotic and early macromolecular synthesis through interaction with light in the planetary context as selective in the evolution of molecules specific for phototrophy.
- Biochemical inheritance: the reuse or modification of molecular components and metabolic pathways that evolved earlier.
- Environmental inheritance and niche construction: the biogeochemical and climatic modification of planetary environments by phototrophs, which changes the conditions for subsequent evolution.
- Ecological priority effects: the occupation of ecological niches by established phototrophic systems, potentially limiting the establishment of alternative strategies.
- Planetary disturbances: large-scale environmental changes (e.g., Snowball Earth Event and GOE) that disrupt existing ecosystems, relax earlier constraints, and create new evolutionary opportunities.

## 7.2. Extent of a photosynthetic biosphere over the surface of a planet

Contributors: Nancy Kiang, Greg Fournier, Tejinder Singh

**Q7.2: What controls the proliferation/productivity of phototrophy over a planet's surface?**

Organisms are adapted to environmental niches that are heterogeneously distributed over the surface of a planet: chemical (electron donors and acceptors, pH, nutrients), physical (temperature, water, radiation/UV), ecological (competition, mutualism, disturbance). Using Earth history as a series of analogs for the origin and spread of phototrophs, it is clear these conditions are potentially highly variable across inhabited worlds. In biosignature detectability studies, prescribing maps of climate-biome associations has been the necessary first step to introduce surface life to test case scenarios. Thus far, these have included using the Earth-through-time as a model to conjecture about potential dense distribution of anoxygenic phototrophs in the Archean oceans, or mixed community microbial mats confined to shallow coastal zones in the Proterozoic, and ecosystem types maps for the modern Earth (HWO25-SBTG), and the change in strength in the VRE with changing continents (Kaltenegger et al. 2007), and changing vegetation with Quaternary climate extremes (Arnold 2008).

Future work will need to explore other scenarios of origins and spread between land, aquatic, and marine environments; extremophiles (see environmental ranges in Merino et al. (2019)), and, in general, a full diversity of biota in diverse planetary scenarios, important variables for which are summarized in the next Section 8 Parameter space for other planets. Phototrophic organisms may have originated in marine hydrothermal vent environments or freshwater hot springs and serpentinizing systems on land (Martin et al., 2018; Nisbet et al., 1995), while phylogenetic reconstruction strongly points to a freshwater origin of oxygenic photosynthesis (Sánchez-Baracaldo & Cardona, 2020; Sánchez-Baracaldo et al., 2014). Plausible phototrophic spectral features would have to be estimated, as was summarized in Section 5.2.1, and potential productivity for different evolutionary stages, and in addition their density. Perhaps most challenging and unsolved is that rocky exoplanets have currently no theoretical constraints on their surface composition of continent vs. ocean, surface or ocean chemistry, or even atmospheric pressure. Rather than devising narrowly tailored scenarios, it will be necessary to devise techniques to generate perturbed parameter ensembles of planetary surface and atmospheric compositions to expand the libraries of potential planetary spectra.

Methods to investigate the potential surface distribution of environmental niches for photosynthesis on other planets include modeling of 2D or 3D climatology taking into account numerous planetary parameters and feedbacks between the biosphere and atmosphere (Section 8), the capabilities of which are reviewed in Schwieterman et al (2018). For surface chemistry and availability of electron donors, unfortunately, a predictive theory or method is not yet available. The spatial heterogeneity of ecosystems is both an emergent property of adaptations to the environment as well as a determinant of their radiative transfer and spectral reflectance as observed from space. As an alternative to the complexity, large uncertainties, empirically parameterized dynamic global vegetation simulation models (DGVMs) used in Earth Science to predict vegetation cover and activity over the Earth, astrobiology research for phototrophic signatures should develop generalized scaling theories to predict distributions of biophysical features of biota into chemical and climate niches on a planet, e.g. Kempes et al. (2011); (Kottek et al., 2006).

Advancing our ability to constrain the surface distribution of phototrophic biota over diverse environmental niches on a planet will require collaborations among biophysicists and microbiologists to determine likely spectral features of phototrophy, ecophysiologists identify likely physiological limits (Section 9.2), theoretical ecologists to develop scaling theories for productive metabolisms and organism morphology and density or biomass in different environments, and possibly Earth System Modelers to validate scaling theories coupled to atmospheric models utilizing the Earth satellite record.

## 8. Parameter space for other planetary environments

Contributors: Eddie Schwieterman, Vikki Meadows, Avi Mandell, Karl Stapelfeldt, Nancy Kiang, Niki Parenteau, Massimo Maffei, Tejinder Singh

**Q8a: What planetary parameters control the surface environment to support phototrophy ?**

**Q8b: How do we demarcate the photosynthetic "habitable zone" around a star?**

Devising search strategies for exoplanet-observing missions to detect signs of phototrophy to date has relied on evolving understanding of what defines the habitable zone around a star and instrument designs with the spectral bands and resolution to resolve the known features of light harvesting pigments on Earth. Previous modeling studies have explored for oxygenic photosynthesis the outer edge limits for both surface temperature and light from other stellar types (Lehmer et al. 2018; Hall et al. 2023). These studies have been based on simple empirical light limits from select Earth organisms.

Most efforts to date to predict surface photosynthetic biosignatures on habitable-zone planets orbiting other stars have focused on photopigment optimization for the time-averaged stellar spectrum (Kiang et al., 2007; Lingam et al., 2021; Lehmer et al., 2021; Duffy et al., 2023), treated in Section 5.2.1 Spectral peaks and bands. There has been emerging research surrounding the potential for prebiotic chemistry based on the time-averaged UV stellar spectrum and predicted surface UV environment, leading to the concept of an "abiogenesis zone" with possible limitations for M dwarf planets (Ranjan et al., 2017; Rimmer et al., 2018, 2021). Not yet explored are how the stellar type, planet formation, and composition with distance from a star would affect the prebiotic chemical conditions for synthesis of molecules relevant to the origin of photosynthesis.

Common physical and chemical environmental variables that are used to categorize extremophiles on Earth include temperature, pressure, pH, salinity, and water activity (Merino et al. (2019), https://doi.org/10.3389/fmicb.2019.00780). The extreme ends of these variable ranges generally correspond to low growth and productivity (i.e., the kinetics that drive metabolic reactions are slow), which is particularly true for polyextremophiles, those organisms that can persist in the face of multiple environmental extremes (Capece et al., 2013; Harrison et al., 2013). The environmental tolerances of photosynthesizers and phototrophs are typically more restricted than those of their chemosynthetic and chemotrophic counterparts (e.g., Seckbach and Oren, 2007; Ward et al., 2012). However, adaptations to extremes could manifest as observable features at scales ranging from the individual organism to the planetary ecosystem. It is also crucial to acknowledge that physical and geochemical environments may vary widely on a planetary scale and, as on Earth, change markedly with the planet's biogeochemical evolution (Lyons et al., 2014).

The possible behavior of prebiotic and pre-photosynthetic chemical reactions in whole organisms and ecosystems, and the corresponding observable adaptations, through alteration in the pigments, morphology, and temporal behavior that combined affect the reflectance signature of phototrophs, are relatively understudied across a wide range of parameter space, which can be broken down into **stellar**, **orbital**, **planetary,** and **atmospheric** variables, all of which are potentially observable and quantifiable, depending on system properties and observing mode and fidelity.

**Stellar** variables beyond the time-averaged spectrum include *spectral variability* due to flares, which can differ in *intensity*, *cadence*, and *wavelength-dependent effects*. The *age* and *composition* of the star may likewise hold clues to the composition and tectonic and atmospheric evolution of its orbiting planets, though with many caveats.

**Orbital** parameters include *semi-major axis* and *eccentricity*, which control time-averaged and time-dependent insolation, influence global climate and circulation patterns, and may prescribe limits on the greenhouse forcing required to maintain clement conditions, such as $CO_2$ levels, which will have downstream chemical impacts. In particular, the majority of the Earth analog studies developed to inform missions like HWO assume the planet is at Earth's current position at the the inner edge of the habitable zone, where the carbonate silicate cycle maintains $CO_2$ at relatively low levels of 100s of ppm. At further orbital distances, which is the vast majority of the search phase space for HWO, planets would be predicted to have much higher atmospheric $CO_2$ abundances, up to several bars, and the biological consequences of these environments for the mechanisms and productivity of photosynthesis, and the detectability of their products, is important to consider.

**Planetary** parameters of consequence that are observable or inferable include *radius* and *surface gravity,* both of which can impact climate and circulation patterns, with surface gravity an important consideration for capillary action and the limiting physical sizes for vegetation analogs (Cordier et al., 2024; Dartnell, 2012; Doughty & Wolf, 2010). Surface gravity depends on *mass* and radius, which together determine the *bulk density* and constrain composition (Rogers & Seager, 2010; Unterborn et al., 2023). Other physical planetary parameters include *obliquity* and *rotation rate,* which also impact climate and circulation. Obliquity, along with orbital eccentricity, will modulate seasonality, which may be either damped or more extreme than on Earth, thereby distributing surface life and its attendant adaptations across different climate zones and affecting their potential observable signatures. More subtle planetary parameters include the *ocean/land* fraction, which will affect the environmental conditions and potentially observable features of photosynthetic organisms; and the *magnetic field*, which could influence the evolution and manifestation of photosynthesis, starting from different rates, yields, and product distributions of biochemical reactions and selectivity in prebiotic and early macromolecular biosynthesis conditions (Rodgers, 2009), the evolution of the atmosphere (Tarduno et al., 2025) and primitive phototrophs through to higher plants (Maffei, 2014; Maffei, 2022).

**Atmospheric** properties can profoundly influence the adaptations and observable signatures of photosynthetic organisms. The atmospheric mass, along with the surface gravity, determines the surface *pressure*. High pressure can cause pressure broadening of photosynthetic pigments and a red shift in their wavelength of maximum absorption (Jalviste et al., 2020; Mizoguchi et al., 2008). The thickness (*column density*) of the atmosphere also determines the amount of ionizing radiation from solar energetic particles (SEPs) and galactic cosmic rays (GCR) that can reach the surface. Planets with thin atmospheres, and especially those orbiting closer to their host star, such as M-dwarfs, can be subjected to intense ionizing radiation from SEPs (Airapetian et al., 2020; Yamashiki et al., 2019). The atmospheric pressure and molecular *composition* (especially greenhouse gases) substantially influence the planet's atmospheric structure and climate. The mass and composition of the atmosphere, along with *condensates* (clouds) and *aerosols*, also dictate the atmosphere's transmissivity and the resulting wavelength-dependent surface radiation environment, distinct from the top-of-atmosphere stellar spectrum.

Some **surface environmental** variables, such as *temperature*, are potentially measurable (e.g., by mid-infrared emission spectroscopy) or inferable (from climate modeling combined with the observable parameters above), while chemical surface variables like pH, salinity, and water activity are not directly observable. However, the atmospheric composition, specifically $CO_2$ content, can influence pH, and broad trends within the habitable zone can be inferred, with potential implications for the chemical and evolutionary challenges faced by different forms of life (Schwieterman et al., 2019). Likewise, the example of Venus suggests parameters like water activity may be partly inferable from atmospheric composition (Hallsworth et al., 2021). Surface liquid water may possibly be directly detected by observing glint from an ocean (Lustig-Yaeger et al., 2026; Robinson et al., 2010; Williams & Gaidos, 2008) , or inferred from rainbows and scattering by clouds and atmospheric gases (Vaughan et al., 2023). The wavelength-dependent reflectivity of the planetary surface is seen filtered through the overlying atmosphere. However, it may be possible to perform spectral retrievals that simultaneously fit for the atmosphere and surface properties, and recent studies show that these retrievals of surface properties are improved if cloud optical properties are known, or additional wavelengths are measured (F. Wang et al., 2022). Attributing extant surface chemistry is a current major challenge; while the field of astrochemistry can place constraints on prebiotic composition, an exoplanet's potential stage of surface evolution will not be possible to constrain given current understanding.

The ultimate connection between factors such as habitable zone position, atmospheric composition, and other planetary parameters and the evolutionary potential and spectral manifestation of photosynthetic life, specifically, is underexplored.

Key priorities to address this parameter space are not only to investigate sensitivities of life process along any one axis but devise methods to quantify the interactive effects among multiple parameters, and from molecular scale laboratory and theoretical modeling studies to planetary scale general circulation models. For each of the above variables, there are gaps in studies to cover the parameter space, as well as sometimes model limitations Such studies will require strongly cross-disciplinary collaborations, as discussed further in Section 9.

## 9. Cross-disciplinary and methodological opportunities

What methods can we use to characterize biomolecules at or near their origin? To quantify environmental limits for phototrophs? To discover novel phototrophs on Earth? To detect phototrophy on another planet?

### 9.1. Origins

Contributors: Betül Kaçar, Massimo Olivucci, Keiichi Inoue

**Q9.1: What are methods needed to better constrain origins of molecules relevant to photosynthesis?**

As discussed in earlier sections, photosynthesis is not a single molecular innovation, but relies on the emergence and co-evolution of a suite of interconnected processes involving pigments, cofactors, reaction centers, electron transport chains, carbon fixation pathways, and environmental energy sources. Understanding its origin therefore requires methods that can reconstruct both the molecules themselves and the planetary conditions under which they could have functioned.

As observed in Section 4.1, the fossil evidence for the earliest photosynthesis is poorly constrained prior to 3.0 Ga. The scientific consensus to date is that the earliest life was not photosynthetic, but precursors to its various components possibly derived from prior roles in chemolithoautroophic organisms (Mulkidjanian & Junge, 1997; Schopf, 2021), and anoxygenic photosynthesis preceded oxygenic photosynthesis (Hohman-Marriott & Blankenship, 2011). Sections 4.2-4.5 on the origins of porphyrins and chlorophylls, proton-pumping rhodopsins, and water oxidation all raise the similar issue of whether or how these were chemistry available that was adopted by biology, or were actually biological inventions, possibly having prior functions in earlier organisms that were then linked to phototrophy. Experimental and theoretical studies to evolve forward from primitive precursors would help bridge the gap between origin of life science and the earliest evidence for phototrophy.

Geological evidence such as stromatolites, carbon isotopes, sulfur isotopes, iron formations, redox proxies, and atmospheric oxygenation records can constrain when photosynthetic metabolisms became globally significant, but they often cannot identify the precise molecular systems or the mechanisms involved. Because early microbial life lacks most or all observable characteristics of early biotic molecules, the fossil record provides a poor basis for historical reconstruction based on paleontological data alone for understanding the origins of photosynthesis. It is here that gaps exist between prebiotic chemistry, ancient life, and modern biology, in part due to the inherent limitation of the Precambrian fossil record (Schopf, 2021). Methods have been developed to combine geological and geochemical evidence with information preserved in modern biomolecules, such as genomes and proteins (e.g. Nishihara et al. (2024)) to reconstruct backward in time ancient forms of biomolecules.

Comparative genomics, metagenomics, and functional metagenomics can help identify the diversity and environmental distribution of genes involved in phototrophy. These approaches can be applied to photosystems, chlorophyll and bacteriochlorophyll biosynthesis genes, retinal-based rhodopsins, carbon fixation enzymes, electron transport proteins, and accessory pigments. Homology searches can reveal how widely these genes are distributed, while functional metagenomics can test whether environmental sequences encode light-harvesting, ion-pumping, or redox functions. These methods are especially useful for detecting phototrophic strategies that may not leave obvious fossil signatures.

For example, as applied to rhodopsins, one useful approach is to treat proteins as "molecular fossils." (Sephus et al., 2022). In a previous study, modern metagenomic and functional metagenomic surveys identified rhodopsin-like genes across diverse environments, while homology searches revealed their taxonomic distribution and ecological associations (Béjà et al., 2000; Alina Pushkarev et al., 2018; Rodriguez-Valera et al., 2022), as well as ancestral sequence reconstruction of ancient rhodopsins (for example (Dungan & Chang, 2022) and (Dungan & Chang, 2022) ). These data were then used in phylogenetic and ancestral sequence reconstruction approaches to infer ancient rhodopsin sequences, predict their structures and functions, and test whether ancestral proteins were likely involved in light-driven ion transport or other phototrophic processes.

By linking reconstructed ancient molecular properties, such as pigment absorption maxima or ion-pumping capacity, to models of early Earth or exoplanet environments, we can ask which phototrophic systems would have been functional under different stellar and atmospheric conditions. Thus molecular evolution can inform about ancient ecological niches and broader questions of habitability.

Constraining the origins of photosynthesis-relevant molecules requires collaboration across fields: prebiotic geochemistry and chemical evolution experiments to define plausible sources of pigments, cofactors, and redox-active compounds; comparative and functional metagenomics to identify modern diversity and environmental context; ancestral sequence reconstruction and laboratory resurrection via the tools of synthetic biology to test ancient and diverse molecular phenotypes; and habitability modeling to evaluate whether inferred functions were viable under ancient planetary conditions. Together, these methods expand the question from "when did photosynthesis appear?" toward a more mechanistic question: which molecules *could and can* emerge, function, and be selected in specific environmental niches on diverse environments, be it the early Earth or on habitable planets around other stars.

## 9.2. Ecophysiological limits and searching for novel phototrophs

Contributors: Nancy Kiang, Nicoletta La Rocca, Elisabetta Liistro, Anastasia Yanchilina, Yu Komatsu, Colin Gates

### Q9.2: Can we determine a phototrophs reflectance spectrum from its environmental niche?

The search for life requires development of a predictive capability for microbial phototroph adaptations (pigment spectra), activity (gas emissions), and surface spectral reflectance in extraterrestrial environments. To achieve this, research is needed in the following,: 1) in situ field studies to quantify how Earth phototrophs are adapted to their native environment's physicochemical and ecological controls to understand why certain biosignatures may dominate; 2) lab experiments to identify biogeochemical products, reaction pathways, and associated diagnostic spectral signatures with phototrophic metabolisms to characterize their optima and limits and to investigate their behavior under a diverse range of planetary environments; 3) modeling, synthetic biology, and metagnomic studies to probe theoretical limits for alternative organism adaptations to other environments and to uncover novel phototrophs to expand the database of diversity and that could fulfill predictions of alternative adaptations.

**Field characterization.** For microbial and microalgal phototrophs, it is not common practice to measure or monitor at length the native environmental niche in detail, although this has been done for the spectral light environment of microbial mats (Fenchel & Kuhl, 2000; Pierson et al., 1987) and for *Acaryochloris marina* (Larkum & Kuhl, 2005), for which also activity levels in situ were measured (Kühl et al., 2005). More such field characterization would be desirable for a greater diversity of phototrophs, including organisms with rhodopsin-based proton pumping.

**Lab characterization.** Example lab experiments to characterize the photosynthetic viability and efficiency of different oxygenic phototrophs under an M dwarf simulated spectrum were recently conducted by (Battistuzzi et al., 2023; Battistuzzi et al., 2024; Liistro et al., 2026). Among their observations were that a far-red cyanobacteria required no acclimation to the M dwarf light and showed reflectance spectra demonstrating a NIR-shifted red edge biosignature.

For rhodopsins, light-driven ion-transport activities can be investigated by monitoring pH change in the external medium surrounding bacterial cells (Inoue et al., 2015), by electrophysiological patch-clamp measurements(Okuyama et al., 2024), and by intracellular pH imaging using genetically encoded pH indicators (Nakao et al., 2022). To gain a physicochemical understanding of the photoreaction dynamics of ion-pumping rhodopsins, time-resolved spectroscopic methods can be used to estimate the turnover rates of rhodopsins (Chizhov et al., 1996), while also providing insights into structural and protonation changes that occur during the photoreaction (Smith et al., 1985) (Lórenz-Fonfría & Kandori, 2009). Furthermore, time-resolved serial-femtosecond X-ray crystallography and time-resolved serial synchrotron crystallography can reveal three-dimensional structural dynamics at atomic resolution.

**Discovering novel photosystems or organisms.** While various models have been put forth to predict the absorbance spectra of photosynthetic pigments (Section #.#), none yet is accepted as fully generalizable and quantitative. For rhodopsins, the methods of modern theoretical and computational photochemistry and photobiology, especially multiscale methods, allow investigation and prediction of the adaptation of rhodopsins to light color and light intensity. This is done via prediction of the absorption spectrum, isomerization quantum efficiency and thermal isomerization barrier (Schnedermann et al., 2018). The prediction of vibrational spectra also appears of basic importance as these guarantee a more accurate fingerprint for the presence of a retinal chromophore (Broser et al., 2023). Theories of optimization of photosystems to extraterrestrial conditions can be tested by designing artificial photosynthetic systems for those conditions and evaluating their efficacy through computer simulation, chemical engineering, and synthetic biology approaches. Artificial photosynthesis can in theory far exceed natural limitations (Z. Wang et al., 2022), but further research is needed to optimize type and relative positions of photoabsorber, oxidizing catalyst, electron carrier(s), and product-forming catalyst.

In addition to biophysical methods and enrichment cultures to uncover novel phototrophs, (meta)genomics could supply additional hints on the presence of phototroph genes in a given environmental sample (Antonaru et al., 2020). This could be via homology searches (Béjà et al., 2000) or using functional metagenomics (Alina Pushkarev et al., 2018).

Developing a predictive capability for phototroph biosignatures (spectral reflectance, activity) is needed through: filling the gaps in quantitatively characterizing the environmental niches of phototrophs and their physiological sensitivities and ecological interactions in both native and simulated alternative planetary conditions; theoretical exploration of photosystems adapted to these other environments through modeling and synthetic biology approaches; and further discovery of novel phototrophy through those synthetic approaches as well as metagenomic sampling. This research will require both traditional disciplinary methods applied in different ways, as well as cutting edge lab, modeling, and synthetic biology approaches.

## 10. Detection of phototrophy

Contributors: Anastasia Yanchilina, Mike Wong, Emilie Lafleche, Anna Grace Ulses, Ligia F. Coelho, Yu Komatsu, Vikki Meadows, Yuka Fujii

**Q10a: What suite of observations would support inference of the presence of phototrophy?**

**Q10b: What abiotic background characterization is needed to rule out false positives and false negatives for phototrophy?**

**Q10c: What tools need community support to advance technologies for phototrophy detection?**

Detection of phototrophy in Earth's fossil record is vital for clarifying its origins and evolution on Earth to provide principles for the same on other planets, as was treated in Section 4.1 Geological record. On exoplanets, if extant phototrophic pigments, associated gases, and/or surface reflectance features can be identified remotely, these features—taken together and properly interpreted alongside environmental contextual knowledge of a specific exoplanetary system—might provide a convincing signal of a phototrophic extraterrestrial biosphere by offering alternative lines of evidence for for the same biological phenomena. This need has been outlined in several science cases for future exoplanet direct imaging missions, including NASA's upcoming Habitable Worlds Observatory (HWO) (Parenteau et al., 2026 [and other HWO Living Worlds SCDDs]). Imputing surface spectral features to phototrophy must ultimately account for such biosignatures being inextricable from their environmental context, not only their adaptation to the surface incident light spectral quality and quantity but also other planetary variables outlined in Section 8 Parameter space for other planets.

Recent progress in observation simulation studies have emphasized statistical rigor, retrieval degeneracies, and the need to test abiotic pathways and instrument limits before interpreting any single molecule as evidence of life ( Parkinson et al., 2026)(Meadows et al., 2022; Schwieterman et al., 2018; Schwieterman & Leung, 2024). The primary types of surface signatures of phototrophy for which exoplanet missions could probe and how they might vary were summarized in Section 3.2.2 Spectral peaks and bands, including the VRE and the diversity of light harvesting pigments and photoprotective pigments. Other recent ideas to aid photosynthesis biosignature detection include joint detection of a putative pigment's absorbance feature and its fluorescence to strengthen validation of phototrophy (Frankenberg et al., 2014; Komatsu et al., 2023) Sun et al., 2018). Also, surface biosignatures need not be limited to continents: floating aquatic vegetation can produce VRE features over water-covered surfaces (Murakami et al., 2025), and the Eocene Arctic *Azolla* hypothesis (Brinkhuis et al., (2006), Neville et al., (2019)) offers a possible Earth-based analogue for extensive floating freshwater vegetation over episodically freshened ocean surfaces.

From a detectability standpoint, the atmospheres of Ocean Worlds and Hycean like planets may be among the most accessible laboratories for near-term biosignature searches (Madhusudhan et al., 2021; Mitchell & Madhusudhan, 2025; Seager et al., 2025) (Seager et al., 2025), as their extended atmospheres and large surface reservoirs improve prospects for transmission spectroscopy. It has been suggested that recent JWST transmission spectroscopy of a sub-Neptune reveals the presence of CH4 and CO2, a potential biosignature for terrestrial environments (e.g. (Krissansen-Totton et al., 2018) in addition to DMS/DMDS (Madhusudhan et al., 2025; Madhusudhan et al., 2023), a gas that on Earth, in a terrestrial environment, is predominantly formed by marine phytoplankton. However, the statistical significance of the detection of DMS, and its interpretation, has been called into question by the community (c.f. Stevenson et al., 2025, and references therein). Inference of extraterrestrial phototrophy from planetary spectral signatures given upcoming telescope technology also has unique retrieval challenges. There is a need to characterize more scenarios for potential atmospheric spectral patterns linked to phototrophy: besides the canonical $CH_4/O_2$ chemical disequilibrium there are currently uncharacterized and numerous volatile organic carbons that could affect the complexity of atmospheric composition (Krissansen-Totton et al., 2016; Seager et al., 2016; Walker et al., 2018; Young et al., 2024). A persistent uncertainty is whether biogenic gas signatures associated with phototrophy can survive transport and photochemical loss/alteration sufficiently to be detectable and recognizeable, with the risk of false negatives, an issue raised in OWL2023 but still requiring quantitative constraints.

Recent progress has been made in building tools to better interpret phototrophic signatures in an observational context. Spectral libraries for biological pigments are expanding to include the diversity of phototrophic biosignatures from different environments (Baldridge et al., 2009; Coelho et al., 2022; Meerdink et al., 2016; Niedzwiedzki et al., 2026). Global hyperspectral and temporal surface observations from Earth-observing satellites and their endmember spectral databases have begun returning data (Caplan & Huemmrich, 2025; Green et al., 2023; Lamour et al., 2026). Observation simulation modeling paired with planetary general circulation model (GCM) simulations are increasingly emerging to quantify detectability of features of interest utilizing realistic, multiple-scenario atmospheric and orbital states (Fauchez et al., 2022; Guzewich et al., 2020; Laflèche et al., 2026; Metz et al., 2024). Machine learning approaches are now being used to explore and classify noisy spectra to tease out robust signals in exoplanet spectra (Duque-Castaño et al., 2025).

Future research should prioritize abiotic laboratory and planetary model simulation studies to establish or rule out pathways for phototrophy false-positives. The astrobiology community should seek to collaborate with science teams of recent hyperspectral Earth-observing missions (Earth Surface Mineral Dust Source Investigation, EMIT; Plankton, Aerosol, Cloud, ocean Ecosystem, PACE) in amassing spectral libraries of end member optical properties (ecosis.org) that promise to be a valuable resource. Observation simulations can advance through development of modeling packages that speed the incorporation of spectral libraries and more realistic surface representations directly into forward models and inversion pipelines to quantify more rigorously the detectability of pigment spectral features and phototrophy-linked gases given the impacts of clouds, atmospheric composition, viewing geometry, and signal-to-noise limitations. The simulations of GCM planetary states for input to observation simulations similarly can advance with greater self-consistency of the surface-atmosphere-chemistry-stellar coupling. Moreover, AI/ML models trained on remote sensing data hold promise to advance comparative planetology to classify planets based on their reflectance spectra, identify robust spectral features, and develop probability models to readily distinguish biological signals from abiotic signals.

## 11. Conclusion and recommendations

This white paper fleshes out specific astrobiology questions for photosynthesis research that were only broadly touched upon or missed in previous Astrobiology strategy reports, NASA2015 and OWL2023. In Table #, we list the questions about photosynthesis laid out in NASA2015, next to our refinements to those questions and addition of critical questions that were not previously raised. NASA2015 had asked, “Was phototrophy adapted from simpler forms of light-driven energy transduction, UV protection mechanisms, trophic mechanisms, or some other process?” Little has been done to advance this question since then, so it remains. NASA2015 did not include carbon fixation origins or efficiencies among its photosynthesis questions, but we deemed it important for addressing the large impact of autotrophy on a planet. Hence we also raised the question of rhodopsin-based autotrophy. Since direct imaging technology has been further developed since NASA2015 and the Habitable Worlds Observatory is now a central community endeavor, this white paper includes a summary of biosignature detection issues.

In Table 2, our questions fit easily into the framework of the current NASA-DARES focus areas (FA). At this time, the NASA-DARES FA1 on Origin of Life places heavy emphasis on polymers and information but makes no mention of enzymes or energy acquisition, and the lists of environmental conditions are very much through the lens of chemistry and chemotrophy and neglect biophysics and autotrophy. One of the research areas for FA1 is: “3. From the origin of life to the Last Universal Common Ancestor (LUCA), a. The OoL-LUCA Gap: How did life evolve from its origin to the complexity of LUCA across the missing evolutionary interval?” Since photosynthetic organisms most likely came after the first prokaryote, their origin occurs in the OoL-LUCA Gap. Overall, the DARES draft strategy provides an encompassing framework that can cover research questions about the origin, evolution, proliferation, and detectability of phototrophy, but the specific questions we flesh out here about phototrophy are important for the program managers to know about, very much because review panels for NASA funding calls must be solicited with the appropriate expertise to evaluate phototrophy proposals.

This white paper has highlighted critical questions for researchers to answer to advance the search for phototrophic biosignatures to a predictive capability in collaborations across the wide array of scientific disciplines that provide the astrobiological context. These questions pertain to the origins, key features, diversity, and potential for alternative adaptations in fundamental components of phototrophy: chlorophylls, rhodopsin-based proton-pumping, and light harvesting; the electron transfer pathway in photosynthesis; Rubisco and carbon fixation pathways. We also identify research needs to constrain how these molecular mechanisms evolve and compete, which affects how they scale through time over a planet to have the potential to be detected by a direct imaging mission.

As with all astrobiology, the endeavor to constrain phototrophic biosignatures as a co-evolved adaptation to their environment is necessarily highly cross-disciplinary. Specifically for phototropic biosignatures, essential cross-disciplinary collaborations include:

- Astronomers, planetary scientists, geochemists, atmospheric scientists, oceanographers, and climate modelers to determine spectral irradiance reaching a planetary surface and under water, the potential electron donor resources, and the variability of the planet's surface habitability for phototrophy.
- Biophysicists, biochemists, structural and theoretical chemists, and synthetic biologists to quantify the efficiency trade-offs in adaptive variations in the photosynthetic apparatus and in rhodopsin-based proton-pumping.
- Inorganic, organic, structural chemists and theoretical chemists/physicists to determine the limits on photochemistry and charge separation, as well as the development of the underlying biochemistry in response to planetary availability of metals and redox donors/acceptors.
- Synthetic with computational chemists and biologists, bioinformaticians, microbiologists, plant ecophysiologists to identify what adaptations have or can occur under a variety of light spectral quality and quantity.
- Chemists, physicists, microbiologists, plant ecophysiologists, and biometeorologists to discover novel phototrophs on Earth and quantify how alternative planetary parameters will lead to changes in phototroph chemical activity, organism morphology, interaction with the atmosphere, and biosignatures.
- Evolutionary biologists, geobiologists, paleoclimatologists, ecologists, microbiologists, plant ecophysiologists, remote sensing scientists to determine when and how phototrophs originated, diversified, and spread across a planet, and what type of signal will dominate.
- Astronomers, mission specialists, GCM modelers, data scientists, geochemists, biologists of all kinds to model surface and atmospheric states of exoplanets to input to observation simulators for precursors biosignature detectability studies.

These interdisciplinary collaborative efforts will advance the community toward a predictive science to constrain uncertainties in inferring the presence of phototrophic life from direct imaging observations of other habitable worlds. For successful collaborations, we note that Astrobiology requires training outside a scientist's discipline to engage with colleagues in other fields. Also, scientists who are new to Astrobiology benefit from some guidance to get acquainted with the frameworks of this wide-reaching pursuit. To this end, the NExSS Extraterrestrial Photosynthesis Workshop developed two online introductions by experts in their fields to provide this training, one **Primer on Exoplanets and Astrobiology** to introduce photosynthesis researchers unfamiliar with Astrobiology to the current trends shaping the search for life elsewhere, and a second **Primer on Photosynthesis**, to expand awareness for astronomers, planetary scientists, and other interested scientists about the fundamentals of phototrophy and photosynthesis and the beautiful, elegant science around the process that is the foundation for nearly all life on our planet. These primers are available online at:

https://nexss.info/extraterrestrial-photosynthesis-workshop/

Table 1. NASA Astrobiology Strategy 2015 questions and updated photosynthesis questions identified in this White Paper

| NASA Astrobiology Strategy 2015 | This Paper |
|---|---|
| 1. How soon after the origin of life did phototrophy arise? | Q4.1a: What geological, geochemical, and phylogenomic evidence do we have now for early photosynthesis, and what still needs to be constrained?<br>Q4.2: How and when did chlorophyll emerge?<br>Q4.3a: How and when did retinal and rhodopsin emerge?<br>Q4.3b: Could rhodopsin-based proton-pumping have emerged prior to oxygenic photosynthesis?<br>Q4.4: How and when did photochemical reactions centers (RCs) emerge?<br>Q4.5: How and when did water oxidation and the oxygen-evolving complex emerge?<br>Q4.6: How did Rubisco originate?<br>Q5.3b-1: Can photosynthesis occur without membranes in a soluble system?<br>Q9.1: What are methods needed to better constrain origins of molecules relevant to photosynthesis? |
| 2. How does phototrophy impact the way life interacts with the surrounding geochemistry, and to what extent does | Q4.1b: What delayed the rise of atmospheric O2 after the emergence of oxygenic photosynthesis?<br>Q52.1.c: Can we formulate a predictive science for the wavelength of peak absorbance of a phototroph as a whole organism in arbitrary environments? |

| NASA Astrobiology Strategy 2015 | This Paper |
|---|---|
| this relationship provide biosignatures? | Q5.2.3d: What light harvesting capability would be competitive (most productive) in alternative environments?<br>Q5.3b-1: Can photosynthesis occur without membranes in a soluble system?<br>Q3.2.3b-3: What alternative electron carriers are plausible?<br>Q6.1a: Why does the Rubisco-dependent Calvin-Benson-Bassham cycle dominate on Earth as a carbon fixation pathway?<br>Q7.2: What controls the proliferation/productivity of phototrophy over a planet's surface?<br>Q6.3b: Can we constrain the co-evolutionary stage of a planet and biosphere from limited observational data?<br>Q6.1b: What could be the theoretical ultimate potential efficiency of carbon fixation on Earth in the future?<br>Q6.1cc: How may carbon fixation efficiency evolve as a planet's CO2 level is drawn down over geologic time with the carbonate-silicate cycle evolution with stellar evolution?<br>Q6.3a: Can we constrain the continuous photosynthetic habitability lifetime and orbital distance of a fiducial Earth around another star?<br>Q7.1a: On Earth, how have the dominant signatures of photosynthesis over the course of co-evolution with the planet relied on selectivity for early biomolecules and earlier successional types of organisms?<br>Q7.1b: On another planet, what other paths could lead to the same outcome or diverge to other dominant signatures at different stages of a planet's evolution?<br>Q7.1c: Does photosynthesis inevitably evolve to become more oxidizing?<br>Q7.1d: How likely should it be for a planet to be dominated by a few photopigments versus exhibit diverse photopigments with more complex biosignatures?<br>Q9.2: Can we determine a phototroph's reflectance spectrum from its environmental niche? |
| 3. What did the earliest forms of phototrophy look like? | Q4.2: How and when did chlorophyll emerge?<br>Q4.3a: How and when did retinal and rhodopsin emerge?<br>Q4.3b: Could rhodopsin-based proton-pumping have emerged prior to oxygenic photosynthesis?<br>Q4.4: How and when did photochemical reactions centers (RCs) emerge? |

| NASA Astrobiology Strategy 2015 | This Paper |
|---|---|
| | Q4.5: How and when did water oxidation and the oxygen-evolving complex emerge?<br>Q4.6: How did Rubisco originate?<br>Q5.3b-1: Can photosynthesis occur without membranes in a soluble system? |
| 4. What photoreactive chemicals were available in the environment of early life? | Q4.2: How and when did chlorophyll emerge?<br>Q4.3a: How and when did retinal and rhodopsin emerge?<br>Q4.3b: Could rhodopsin-based proton-pumping have emerged prior to oxygenic photosynthesis?<br>Q4.4: How and when did photochemical reactions centers (RCs) emerge?<br>Q4.5: How and when did water oxidation and the oxygen-evolving complex emerge?<br>Q4.6: How did Rubisco originate?<br>Q5.3c: How does planetary chemistry (i.e. trace metal availability, geochemistry, or availability of electron donors and acceptors) influence the evolution of photosynthetic electron transfer? |
| 5. Was phototrophy adapted from simpler forms of light-driven energy transduction, UV protection mechanisms, trophic mechanisms, or some other process? | Q4.2: How and when did chlorophyll emerge?<br>Q4.3a: How and when did retinal and rhodopsin emerge?<br>Q4.3b: Could rhodopsin-based proton-pumping have emerged prior to oxygenic photosynthesis?<br>Q4.4: How and when did photochemical reactions centers (RCs) emerge?<br>Q4.5: How and when did water oxidation and the oxygen-evolving complex emerge?<br>Q4.6: How did Rubisco originate?<br>Q5.1: Is chlorophyll ultimately the best molecule for light harvesting and charge separation?<br>Q5.3c: How does planetary chemistry (i.e. trace metal availability, geochemistry, or availability of electron donors and acceptors) influence the evolution of photosynthetic electron transfer? |
| 6. When did organisms evolve oxygenic photosynthesis, what were its evolutionary precursors, and how is this innovation related to the general oxidation of the Earth's crust and atmosphere? | Q4.2: How and when did chlorophyll emerge?<br>Q4.3a: How and when did retinal and rhodopsin emerge?<br>Q4.3b: Could rhodopsin-based proton-pumping have emerged prior to oxygenic photosynthesis?<br>Q4.4: How and when did photochemical reactions centers (RCs) emerge?<br>Q4.5: How and when did water oxidation and the oxygen-evolving complex emerge? |

| NASA Astrobiology Strategy 2015 | This Paper |
|---|---|
| | |
| 7. How much biologically useful energy is produced by bacteriorhodopsin and other photoreactive biochemistries? Is this enough to sustain an organism? | Q6.2a: What limits rhodopsin-based phototrophy from supporting fixation of CO2?<br>Q6.2b: Can rhodopsin-based phototrophy be used to support any of the known carbon fixation pathways or a theoretical construct?<br>Q6.2c: Will rhodopsin-based phototrophy inherently be low productivity and therefore low detectability? |
| 8. Are there undiscovered types of phototrophy present on Earth today? | Q5.3b: What electron-transfer architectures beyond those found on Earth are physically and biologically feasible?<br>Q5.3b-2: Can photosynthesis occur without energy storage that depends on generating proton gradients?<br>Q3.2.3b-3: What alternative electron carriers are plausible?<br>Q6.1d: What alternative carbon fixation schemes could be successful and become prevalent on another planet? |
| 9. What types of light can be biologically useful and how does the range of phototrophy constrain our concept of habitable planets? | Q5.1: Is chlorophyll ultimately the best molecule for light harvesting and charge separation?<br>Q5.2.1a: What is the long wavelength limit for oxygenic photosynthesis or photosynthesis at all?<br>Q5.2.1b: What are the spectral limits to rhodopsin-based proton pumping?<br>Q52.1.c: Can we formulate a predictive science for the wavelength of peak absorbance of a phototroph as a whole organism in arbitrary environments?<br>Q5.2.3a: What are the trade-offs of adaptation or acclimation to alternative photon wavelengths?<br>Q5.2.3c: Can light harvesting pigment spectral tuning to the spectral light environment be considered an agnostic biosignature?<br>Q5.2.3d: What light harvesting capability would be competitive (most productive) in alternative environments?<br>Q5.3a: How is coupling of the electron transfer chain (ETC) and the generation of the proton gradient for ATP production affected by differences in both light quality and quantity?<br>Q5.3a-1: What tradeoffs exist among energetic efficiency, metabolic flux, photoprotection, spectral coverage, photo- |

| NASA Astrobiology Strategy 2015 | This Paper |
| --- | --- |
| | excited state energies of pigments, and the number of photosystems in an electron-transfer pathway?<br>Q5.3b: What electron-transfer architectures beyond those found on Earth are physically and biologically feasible?<br>Q5.3a-1: What tradeoffs exist among energetic efficiency, metabolic flux, photoprotection, spectral coverage, photo-excited state energies of pigments, and the number of photosystems in an electron-transfer pathway?<br>Q7.2: What controls the proliferation/productivity of phototrophy over a planet's surface?<br>Q8a: What planetary parameters control the surface environment to support phototrophy ?<br>Q8b: How do we demarcate the photosynthetic "habitable zone" around a star? |
| 10. How will the biosignatures of pigments differ for different atmospheric compositions and parent stellar type? | Q52.1.c: Can we formulate a predictive science for the wavelength of peak absorbance of a phototroph as a whole organism in arbitrary environments?<br>Q5.2.3a: What are the trade-offs of adaptation or acclimation to alternative photon wavelengths?<br>Q5.2.3c: Can light harvesting pigment spectral tuning to the spectral light environment be considered an agnostic biosignature?<br>Q5.2.3d: What light harvesting capability would be competitive (most productive) in alternative environments?<br>Q5.3a: How is coupling of the electron transfer chain (ETC) and the generation of the proton gradient for ATP production affected by differences in both light quality and quantity?<br>Q5.3a-1: What tradeoffs exist among energetic efficiency, metabolic flux, photoprotection, spectral coverage, photo-excited state energies of pigments, and the number of photosystems in an electron-transfer pathway?<br>Q5.3b: What electron-transfer architectures beyond those found on Earth are physically and biologically feasible?<br>Q5.3a-1: What tradeoffs exist among energetic efficiency, metabolic flux, photoprotection, spectral coverage, photo-excited state energies of pigments, and the number of photosystems in an electron-transfer pathway?<br>Q5.3c: How does planetary chemistry (i.e. trace metal availability, geochemistry, or availability of electron donors and acceptors) influence the evolution of photosynthetic electron transfer? |

| NASA Astrobiology Strategy 2015 | This Paper |
|---|---|
| | Q:5.4 How do compensation strategies for dealing with liabilities of light harvesting affect efficiency, e.g. non-photochemical quenching, coping with oxygen?<br>Q7.1a: On Earth, how have the dominant signatures of photosynthesis over the course of co-evolution with the planet relied on selectivity for early biomolecules and earlier successional types of organisms?<br>Q7.1b: On another planet, what other paths could lead to the same outcome or diverge to other dominant signatures at different stages of a planet's evolution?<br>Q7.1c: Does photosynthesis inevitably evolve to become more oxidizing?<br>Q7.1d: How likely should it be for a planet to be dominated by a few photopigments versus exhibit diverse photopigments with more complex biosignatures?<br>Q7.2: What controls the proliferation/productivity of phototrophy over a planet's surface?<br>Q8a: What planetary parameters control the surface environment to support phototrophy ?<br>Q8b: How do we demarcate the photosynthetic "habitable zone" around a star?<br>Q9.2: Can we determine a phototroph's reflectance spectrum from its environmental niche?d |
| <u>Missing:</u><br>Efficiency, Competition, Ecology | Q5.3b-2: Can photosynthesis occur without energy storage that depends on generating proton gradients?<br>Q:5.4 How do compensation strategies for dealing with liabilities of light harvesting affect efficiency, e.g. non-photochemical quenching, coping with oxygen? |
| Carbon fixation | Q6.1a: Why does the Rubisco-dependent Calvin-Benson-Bassham cycle dominate on Earth as a carbon fixation pathway?<br>Q6.1b: What could be the theoretical ultimate potential efficiency of carbon fixation on Earth in the future?<br>Q6.1cc: How may carbon fixation efficiency evolve as a planet's CO2 level is drawn down over geologic time with the carbonate-silicate cycle evolution with stellar evolution?<br>Q6.1d: What alternative carbon fixation schemes could be successful and become prevalent on another planet? |
| Detectability | Q10a: What suite of observations would support inference of the presence of phototrophy?<br>Q10b: What abiotic background characterization is needed to rule out false positives and false negatives for phototrophy? |

| NASA Astrobiology Strategy 2015 | This Paper |
|---|---|
| | Q10c: What tools need community support to advance technologies for phototrophy detection? |
| Methods | Q9.1: What are methods needed to better constrain origins of molecules relevant to photosynthesis? |

Table 2. NASA Decadal Astrobiology Research and Exploration Strategy (NASA-DARES) Focus Areas and specific photosynthesis research questions identified in this white paper.

| NASA-DARES Focus Areas | This white paper |
|---|---|
| FA1. Protometabolism and Synthesis/Function of Macromolecules in Planetary Environments (Origin of Life ) | Q4.1a: What geological, geochemical, and phylogenomic evidence do we have now for early photosynthesis, and what still needs to be constrained?<br>Q4.1b: What delayed the rise of atmospheric O2 after the emergence of oxygenic photosynthesis?<br>Q4.2: How and when did chlorophyll emerge?<br>Q4.3a: How and when did retinal and rhodopsin emerge?<br>Q4.3b: Could rhodopsin-based proton-pumping have emerged prior to oxygenic photosynthesis?<br>Q4.4: How and when did photochemical reactions centers (RCs) emerge?<br>Q4.5: How and when did water oxidation and the oxygen-evolving complex emerge?<br>Q4.6: How did Rubisco originate?<br>Q5.1: Is chlorophyll ultimately the best molecule for light harvesting and charge separation?<br>Q5.3a: How is coupling of the electron transfer chain (ETC) and the generation of the proton gradient for ATP production affected by differences in both light quality and quantity? Q5.3a-1: What tradeoffs exist among energetic efficiency, metabolic flux, photoprotection, spectral coverage, photo-excited state energies of pigments, and the number of photosystems in an electron-transfer pathway?<br>Q5.3b: What electron-transfer architectures beyond those found on Earth are physically and biologically feasible? |

| NASA-DARES Focus Areas | This white paper |
|---|---|
| | Q5.3b-1: Can photosynthesis occur without membranes in a soluble system?<br>Q5.3b-2: Can photosynthesis occur without energy storage that depends on generating proton gradients?<br>Q3.2.3b-3: What alternative electron carriers are plausible?<br>Q6.2a: What limits rhodopsin-based phototrophy from supporting fixation of CO2?<br>Q6.2b: Can rhodopsin-based phototrophy be used to support any of the known carbon fixation pathways or a theoretical construct?<br>Q6.2c: Will rhodopsin-based phototrophy inherently be low productivity and therefore low detectability?<br>Q9.1: What are methods needed to better constrain origins of molecules relevant to photosynthesis? |
| FA2. Abiotic Organic Production and Chemical Evolution within Planetary Environments | Q4.2: How and when did chlorophyll emerge?<br>Q4.3a: How and when did retinal and rhodopsin emerge?<br>Q4.5: How and when did water oxidation and the oxygen-evolving complex emerge?<br>Q9.1: What are methods needed to better constrain origins of molecules relevant to photosynthesis? |
| FA3. Co-Evolution of Biospheres, Worlds, and Planetary Systems | Q4.3b: Could rhodopsin-based proton-pumping have emerged prior to oxygenic photosynthesis?<br>Q4.5: How and when did water oxidation and the oxygen-evolving complex emerge?<br>Q5.2.3a: What are the trade-offs of adaptation or acclimation to alternative photon wavelengths?<br>Q5.2.3b: What are the trade-offs of adaptation or acclimation to low/high light?<br>Q5.2.3c: Can light harvesting pigment spectral tuning to the spectral light environment be considered an agnostic biosignature?<br>Q6.1a: Why does the Rubisco-dependent Calvin-Benson-Bassham cycle dominate on Earth as a carbon fixation pathway?<br>Q6.1b: What could be the theoretical ultimate potential efficiency of carbon fixation on Earth in the future?<br>Q6.1cc: How may carbon fixation efficiency evolve as a planet's CO2 level is drawn down over geologic time with the carbonate-silicate cycle evolution with stellar evolution?<br>Q6.1d: What alternative carbon fixation schemes could be successful and become prevalent on another planet? |

| NASA-DARES Focus Areas | This white paper |
|---|---|
| | Q5.2.3d: What light harvesting capability would be competitive (most productive) in alternative environments?<br>Q5.3c: How does planetary chemistry (i.e. trace metal availability, geochemistry, or availability of electron donors and acceptors) influence the evolution of photosynthetic electron transfer?<br>Q:5.4 How do compensation strategies for dealing with liabilities of light harvesting affect efficiency, e.g. non-photochemical quenching, coping with oxygen?<br>Q6.3a: Can we constrain the continuous photosynthetic habitability lifetime and orbital distance of a fiducial Earth around another star?<br>Q6.3b: Can we constrain the co-evolutionary stage of a planet and biosphere from limited observational data?<br>Q7.1a: On Earth, how have the dominant signatures of photosynthesis over the course of co-evolution with the planet relied on selectivity for early biomolecules and earlier successional types of organisms?<br>Q7.1b: On another planet, what other paths could lead to the same outcome or diverge to other dominant signatures at different stages of a planet's evolution?<br>Q7.1c: Does photosynthesis inevitably evolve to become more oxidizing?<br>Q7.1d: How likely should it be for a planet to be dominated by a few photopigments versus exhibit diverse photopigments with more complex biosignatures?<br>Q7.2: What controls the proliferation/productivity of phototrophy over a planet's surface?<br>Q9.2: Can we determine a phototroph's reflectance spectrum from its environmental niche? |
| FA4. Comparative Planetology to Understand Habitability | Q8a: What planetary parameters control the surface environment to support phototrophy ?<br>Q8b: How do we demarcate the photosynthetic "habitable zone" around a star? |
| FA5. Detecting Signs of Living Environments and Living Worlds | Q5.2.1a: What is the long wavelength limit for oxygenic photosynthesis or photosynthesis at all?<br>Q5.2.1b: What are the spectral limits to rhodopsin-based proton pumping?<br>Q52.1.c: Can we formulate a predictive science for the wavelength of peak absorbance of a phototroph as a whole organism in arbitrary environments?<br>Q5.2.2: What are the low and high light flux limits for different kinds of phototrophy? |

| NASA-DARES Focus Areas | This white paper |
|---|---|
| | Q10a: What suite of observations would support inference of the presence of phototrophy?<br>Q10b: What abiotic background characterization is needed to rule out false positives and false negatives for phototrophy?<br>Q10a: What suite of observations would support inference of the presence of phototrophy?<br>Q10b: What abiotic background characterization is needed to rule out false positives and false negatives for phototrophy?<br>Q10c: What tools need community support to advance technologies for phototrophy detection? |
| Focus Area 6: Astrobiology-Focused Mission Approaches and Technology Development | Q10c: What tools need community support to advance technologies for phototrophy detection?<br>Q8b: How do we demarcate the photosynthetic “habitable zone” around a star? |
| Focus Area 7: Investment in Astrobiology Physical and Digital Infrastructure | Q10c: What tools need community support to advance technologies for phototrophy detection? |
| Focus Area 8: Workforce and Early Career Support Focus Area 8: Workforce and Early Career Support | |
| Focus Area 9: Astrobiology in Society | |
| No Focus Area fit | Q:5.4 How do compensation strategies for dealing with liabilities of light harvesting affect efficiency, e.g. non-photochemical quenching, coping with oxygen?<br>Q6.2a: What limits rhodopsin-based phototrophy from supporting fixation of CO2?<br>Q6.2b: Can rhodopsin-based phototrophy be used to support any of the known carbon fixation pathways or a theoretical construct?<br>Q6.2c: Will rhodopsin-based phototrophy inherently be low productivity and therefore low detectability? |